\documentclass[prb,preprint,eqsecnum,showpacs]{revtex4}

\usepackage{amsfonts, amsbsy, amssymb, amsmath} 

\usepackage[pdftex]{graphicx}
\usepackage{color}
\usepackage{float}
\restylefloat{figure}
\newcommand{\rmd}{{\rm d}}

\newcommand{\calM}{\mathcal{M}}
\newcommand{\calP}{\mathcal{P}}
\newcommand{\bbP}{\mathbb{P}}
\newcommand{\bx}{\boldsymbol{x}}

\newcommand{\qb}{\bar{q}}
\newcommand{\pb}{\bar{p}}
\newcommand{\bqb}{\bar{\boldsymbol{q}}}
\newcommand{\bpb}{\bar{\boldsymbol{p}}}

\newcommand{\sbar}{\bar{s}}

\newcommand{\Int}[3]{\int_{#2}^{#3}\rmd {#1} \;}

\begin{document}

% Use the \preprint command to place your local institutional report
% number in the upper righthand corner of the title page in preprint mode.
% Multiple \preprint commands are allowed.
% Use the 'preprintnumbers' class option to override journal defaults
% to display numbers if necessary
%\preprint{}

%Title of paper
\title{Microcanonical rates, gap times,\\ and phase space dividing surfaces}

% repeat the \author .. \affiliation  etc. as needed
% \email, \thanks, \homepage, \altaffiliation all apply to the current
% author. Explanatory text should go in the []'s, actual e-mail
% address or url should go in the {}'s for \email and \homepage.
% Please use the appropriate macro foreach each type of information

% \affiliation command applies to all authors since the last
% \affiliation command. The \affiliation command should follow the
% other information
% \affiliation can be followed by \email, \homepage, \thanks as well.

\author{Gregory S. Ezra}
\email[]{gse1@cornell.edu}
%\homepage[]{Your web page}
%\thanks{}
%\altaffiliation{}
\affiliation{Department of Chemistry and Chemical Biology\\
Baker Laboratory\\
Cornell University\\
Ithaca, NY 14853\\USA}

\author{Holger Waalkens}
\email[]{H.Waalkens@rug.nl}
%\homepage[]{Your web page}
%\thanks{}
%\altaffiliation{}
\affiliation{University of Groningen\\
Institute of Mathematics and Computing Science\\
P.O. Box 407\\
9700 AK Groningen\\
The Netherlands}

\author{Stephen Wiggins}
\email[]{stephen.wiggins@mac.com}
%\homepage[]{Your web page}
%\thanks{}
%\altaffiliation{}
\affiliation{School of Mathematics\\University of Bristol\\Bristol BS8 1TW\\United Kingdom}

%Collaboration name if desired (requires use of superscriptaddress
%option in \documentclass). \noaffiliation is required (may also be
%used with the \author command).
%\collaboration can be followed by \email, \homepage, \thanks as well.
%\collaboration{}
%\noaffiliation

\date{\today}

\begin{abstract}

The general approach to classical unimolecular reaction
rates due to Thiele is revisited in light of recent advances in the
phase space formulation of transition state theory for multidimensional systems.
Key concepts such as the phase
space dividing surface separating reactants from products, the average gap time, and the
volume of phase space associated with reactive 
trajectories, are both rigorously defined and readily computed within the
phase space approach.  We analyze in detail the gap time distribution and associated reactant lifetime
distribution for the isomerization reaction HCN $\rightleftharpoons$ CNH, previously
studied using the methods of phase space transition state theory.
Both algebraic (power law) and exponential decay regimes have been identified.
Statistical estimates of the isomerization rate are compared with the
numerically determined decay rate.  Correcting the RRKM estimate to account 
for the measure of the reactant phase space region
occupied by trapped trajectories results in a drastic overestimate of the isomerization
rate.  Compensating but as yet poorly understood trapping mechanisms in the reactant region 
serve to slow the escape rate sufficiently that the uncorrected
RRKM estimate turns out to be reasonably accurate, at least at the particular energy
studied.  Examination of the decay properties 
of subsensembles of trajectories that exit the HCN well through either of 2 available 
symmetry related product channels shows that 
the complete trajectory ensemble effectively attains the full symmetry of the system phase space on
a short timescale $t \lesssim 0.5$ ps, after which the product 
branching ratio is 1:1, the  ``statistical'' value.
At intermediate times, this statistical product ratio is accompanied by 
nonexponential (nonstatistical) decay.  
We point out close parallels between the dynamical behavior inferred from the 
gap time distribution for HCN and nonstatistical behavior recently identified 
in reactions of some organic molecules.

\end{abstract}

% insert suggested PACS numbers in braces on next line
% insert suggested keywords - APS authors don't need to do this
%\keywords{}

%\maketitle must follow title, authors, abstract, \pacs, and \keywords
\maketitle

% body of paper here - Use proper section commands
% References should be done using the \cite, \ref, and \label commands

\section{Introduction}
\label{sec:intro}

The theory of unimolecular reaction rates, both for dissociative and isomerization 
processes, has been of great interest for
nearly a century \cite{Marcelin15}.  Following the original formulation of the 
statistical RRK and RRKM approaches to calculation of reaction rates 
\cite{Rice27,Kassel28,Kassel28a,Marcus51,Marcus52}, there
has been a vast amount of activity, some of which has been described in 
several generations of texbooks 
\cite{Kassel32,Hinshelwood40,Slater59,Bunker66,Johnston66,Rice67,Laidler69,
Robinson72,Forst73,Nikitin74,Smith80,Beynon84,Pritchard84,Wardlaw88,Gilbert90,Baer96,Forst03,Henriksen08}.  
For concise overviews 
of the historical development of unimolecular rate theory, see Forst \cite{Forst03}, 
Chapter 1, also ref.\ \onlinecite{Pollak05}.

To provide the context for the discussion in the present paper, we highlight some
important and relevant contributions to the subject. (The selection of works cited is necessarily 
quite limited.) The development of Slater's ``new'' dynamical approach 
to classical unimolecular dissociation rates \cite{Slater56,Slater59} was followed by Thiele's general
formulation of the problem, which emphasized the dynamical 
significance of the distribution of gap times 
(see below) \cite{Thiele62}.  As we show below, Thiele's work has proved 
to be remarkably prescient in terms of its identification of the appropriate 
phase space structures involved in unimolecular reaction dynamics. 
Following Thiele's work, a large number of computational (trajectory) investigations of
lifetime/gap time distributions have been undertaken (see, for example, results discussed 
in  \cite{Bunker66,Bunker73,Baer96}).  
Dumont and Brumer \cite{Dumont86,Dumont92} and Dumont \cite{Dumont89,Dumont89a,Dumont92a,Jain93}
have analysed unimolecular reaction rates in terms of the 
gap time distribution, while related work 
has been done on the so-called classical spectral theorem \cite{Brumer80,Pollak81}.

A fundamental assumption of the statistical theory of (classical) unimolecular decay is
that intramolecular vibrational energy distribution (IVR) occurs on a timescale much faster than
that for reaction \cite{Robinson72,Baer96}.
Renewed interest in the problem therefore naturally stemmed from
investigations of the properties of molecular vibrational dynamics, modelled as
nonlinearly coupled anharmonic oscillators, in light of the KAM theorem and the
apparent existence of a threshold energy for onset of
global chaos \cite{Tabor81,Rice81,Brumer81,Brumer88}.
The central role of deterministic chaos itself in determining the validity of statistical
approaches to reaction rates, as well as possible quantum manifestations of
classical nonintegrable behavior \cite{Nordholm75}, has received much attention \cite{Brumer88},
most recently in the context of quantum control \cite{Rice00,Gong05}.  
The possibility of mode-specific chemistry also
stimulated much work on the relation between intramolecular dynamics and 
non-statistical reaction dynamics \cite{Thiele80}.  Several examples of 
non-RRKM effects in thermal reactions of medium-sized organic molecules have been 
found in recent years \cite{Carpenter92,Carpenter03,Carpenter05,Bach06}, while 
a detailed discussion of possible non-RRKM effects in one particular reaction, the unimolecular dissociation 
of $\text{H}_3^{+}$, has been given \cite{Berblinger94}.

Advances in the classical theory of chemical reaction rates have been closely linked
to improvements in our understanding of the phase space structure of Hamiltonian systems
\cite{MacKay87,Lichtenberg92,Wiggins92,Arnold06}.  Both conceptually and
in practice it is important to distinguish between systems with 2 degrees of 
freedom (DoF) and multimode ($N \geq 3$ DoF) systems. 

For 2 DoF systems described by standard kinetic plus potential Hamiltonians, 
periodic orbit dividing surfaces 
(PODS; that is, dividing surfaces in configuration space obtained 
by projection of a periodic orbit from phase space)
were shown to have the property of minimal flux (as a consequence of the principle of stationary
action) and to be (locally) surfaces of no return (the velocity vector
is nowhere tangent to the configuration space projection of the periodic orbit) 
\cite{Pechukas81,Pechukas82,Pollak85}.  In this case the PODS 
therefore provide a rigorous realization of the 
concept of the transition state \cite{Wigner38,Truhlar96}.
(PODS have also been defined for 2 DoF systems lacking time-reversal symmetry \cite{Jaffe00}.)
Correlations between features of the classical phase space structure and the behavior 
of the computed reactive flux \cite{Chandler78,Chandler87}
were explored relatively early on for 2 DoF systems by DeLeon and Berne \cite{DeLeon81,Berne82}. 
Important theoretical advances were subsequently 
made relating non-statistical behavior to
molecular phase space structure \cite{Davis92}, in particular the 
existence of intramolecular bottlenecks 
to energy transfer \cite{Bensimon84,Mackay84,Davis85},
broken separatrices \cite{Channon80,Gray86a,Gray87},  and reactive islands and cylinders 
\cite{Marston89,DeAlmeida90,DeLeon91,DeLeon91a,DeLeon92,DeLeon92a,DeLeon94}. 

The multimode case remains less thoroughly explored and understood.
Methods for defining approximate intramolecular 
bottlenecks and reactive dividing surfaces have been devised 
\cite{Tersigni88,Tersigni90,Gray86,Zhao92,Zhao92a,Jang93,Rice96}.
Although the Arnold web of resonances provides a useful 
framework for mapping and analyzing evolution of phase points in near-integrable 
multimode systems \cite{Lichtenberg92,Toda02}, so-called 
``Arnold diffusion'' \cite{Arnold64,Lichtenberg92} 
has become a convenient but often ill-defined catchall 
term employed to describe a variety of possibly
distinct phase space transport mechanisms in $N \geq 3$ DoF systems \cite{Lochak99}.
The possible role of ``Arnold diffusion''   
in multimode molecular systems has been studied in IVR 
\cite{Heller95,Heller99,Leitner97,Toda02,Keshavamurthy07}
and in isomerization \cite{Toda05,Shojiguchi08}.
Local random matrix theories have 
served as the foundation for quantum theoretical treatments 
of isomerization reactions in large molecules \cite{Schofield94,Gruebele00,Gruebele04,Leitner06,Leitner08}, 
while attention has also been given to related scaling approaches
\cite{Schofield93,Schofield95,Keshavamurthy99} 
and to fractional kinetics \cite{Zaslavsky05,Shojiguchi07,Shojiguchi07a}.

An important advance was the realization that normally hyperbolic invariant manifolds 
(NHIMs \cite{Wiggins90,Wiggins94}) provide a natural and theoretically well-founded
generalization of PODS to the $N \geq 3$ DoF case (cf.\ ref.\ \onlinecite{Pollak78}).  
The stable and unstable manifolds 
associated with NHIMs define codimension-one dividing surfaces \cite{footnote1} 
on the constant energy manifold \cite{Wiggins90,Wiggins94}, and so are possible
candidates for reactive separatrices, while the NHIMs themselves
define phase space transition states \cite{wwju,ujpyw}.
Some early attempts were made to compute and visualize such manifolds 
in a 3 DoF system describing surface diffusion of atoms \cite{Gillilan90} 
and for a 4D symplectic mapping modelling
the dissociation of a van der Waals complex \cite{Gillilan91,Toda95} .

As discussed in more detail below, based on the notion of the NHIM and on the development of 
efficient algorithms for computing normal forms at saddles, there has been
significant recent progress in the development and implementation 
of phase space transition state theory 
\cite{Wiggins90,wwju,ujpyw,WaalkensBurbanksWiggins04,WaalkensWiggins04,WaalkensBurbanksWigginsb04,
WaalkensBurbanksWiggins05,WaalkensBurbanksWiggins05c,SchubertWaalkensWiggins06,WaalkensSchubertWiggins08}
(see also 
\cite{Komatsuzaki00,Komatsuzaki02,Komatsuzaki05,Wiesenfeld03,Wiesenfeld04,Wiesenfeld04a,Gabern05,Gabern06,Shojiguchi08}).

In the present paper we consider the problem of defining and evaluating  theoretical
unimolecular reaction rates in light of the penetrating 
analyses of Thiele \cite{Thiele62}, Dumont and Brumer \cite{Dumont86}, and
DeLeon and Berne \cite{DeLeon81,Berne82},  
the classical spectral theorem \cite{Brumer80,Pollak81}, 
and the recent developments in phase space transition state theory 
\cite{wwju,ujpyw,WaalkensBurbanksWiggins04,WaalkensWiggins04,WaalkensBurbanksWigginsb04,
WaalkensBurbanksWiggins05,WaalkensBurbanksWiggins05c,SchubertWaalkensWiggins06,WaalkensSchubertWiggins08}
mentioned above.  The particular reaction chosen for study is the isomerization 
$\text{HCN} \rightleftharpoons \text{CNH}$
\cite{Gray80,pmpb1,pmpb2,Lan93,Bentley93,Tang94,Shah99,WaalkensBurbanksWigginsb04,WaalkensBurbanksWiggins05,Li05,Gong05a}, 
for which several relevant theoretical quantities
have recently been computed for a (classical) model of the HCN molecule 
at fixed energy \cite{WaalkensBurbanksWigginsb04,WaalkensBurbanksWiggins05}.

We are concerned with the rate of reaction,  at fixed  energy, for a 
system described by a time-independent,  $n$ degree-of-freedom (DOF) classical Hamiltonian.  
One measure of the rate at which trajectories leave a region of the energy surface is 
given by the (magnitude of the) flux of trajectories leaving that region (units of energy surface volume/time) 
divided by the energy surface volume of initial conditions in that region 
{\em corresponding to trajectories that will eventually leave the region}.  
This rate is just the inverse of the mean passage or gap time \cite{Thiele62,Dumont86}.
Hence, to compute this rate at fixed energy one must first  (1) define the region of reactants,  
then (2) compute the flux of trajectories exiting this region and, finally, 
(3) compute the volume of the energy surface corresponding to initial conditions 
of trajectories that leave the reactant region. 

The importance of the gap time distribution was emphasized by Thiele \cite{Thiele62}, 
who explicitly invoked the concept of a phase space dividing surface  separating
reactants and products \cite{Wigner39,Keck67}.
While the use of dividing surfaces  (transition states) defined in configuration space
is quite familiar in the field of reaction rate theory \cite{Anderson95,Truhlar96},
carrying out these steps in phase space, as opposed to configuration space,  
remains less familiar in practice \cite{Davis92,Rice96}. 
In the present paper we show how the 
expression for the microcanonical rate of reaction described above 
can be evaluated using the 
phase space approach to reaction dynamics developed in a recent series of papers 
\cite{wwju,ujpyw,WaalkensBurbanksWiggins04,WaalkensWiggins04,WaalkensBurbanksWigginsb04,
WaalkensBurbanksWiggins05,WaalkensBurbanksWiggins05c,SchubertWaalkensWiggins06,WaalkensSchubertWiggins08}.
Moreover, we analyze in detail the properties of the gap time
distribution previously obtained for HCN isomerization using the phase space reaction rate theory 
\cite{WaalkensBurbanksWiggins05}.

The stucture of the paper is as follows: in Section \ref{sec:unimolecular} 
we outline an approach to the classical theory of unimolecular reaction 
rates based upon the general formulation of Thiele \cite{Thiele62}.
We demonstrate that various dynamical quantities related 
to the unimolecular reaction rate defined within Thiele's approach 
can be rigorously defined and computed within the recently developed 
theoretical framework for phase space transition state theory 
\cite{wwju,ujpyw,WaalkensBurbanksWiggins04,WaalkensWiggins04,WaalkensBurbanksWigginsb04,
WaalkensBurbanksWiggins05,WaalkensBurbanksWiggins05c,SchubertWaalkensWiggins06,WaalkensSchubertWiggins08}.
Special emphasis is placed on the properties of the gap time distribution, and the 
relation between the inverse gap time and the rate of reaction.  
In Section \ref{sec:HCN_data} we consider the dynamics of 
the HCN isomerization reaction.  We analyze previously computed classical trajectory
results \cite{WaalkensBurbanksWigginsb04,WaalkensBurbanksWiggins05} in light of the
discussion given in Section \ref{sec:unimolecular}.  Both gap time and reactant
lifetime distributions are analyzed, and distinct temporal regimes identified, with
power law (algebraic) decay seen at intermediate times and exponential decay at long times. 
Moreover, we find a statistical (1:1) product ratio for both algebraic and exponetial decay regimes.
Comparisons are made between computed values of the decay rate and 
statistical estimates of the isomerization rate.  
Our results show that analysis of gap time distributions computed using phase space dividing surfaces
having minimal flux and no-recrossing properties yields a great deal of 
information about the system dynamics.
Section \ref{sec:summary} concludes.

\newpage

\section{General approach to unimolecular reaction rates}
\label{sec:unimolecular}

In this section we present a general formulation of unimolecular reaction 
rates based upon that originally given by Thiele \cite{Thiele62}.  
In their general form the rate expressions derived by Thiele explicitly
invoke the existence of a \emph{phase space} dividing surface separating
reactants and products; such surfaces, although discussed by Wigner \cite{Wigner39}
at the inception of the theory of chemical reaction rates 
(see also \cite{Keck67,Anderson95}), have only recently become amenable 
to direct computation via the use of normal form 
approaches 
\cite{wwju,ujpyw,WaalkensBurbanksWiggins04,WaalkensWiggins04,WaalkensBurbanksWigginsb04,
WaalkensBurbanksWiggins05,WaalkensBurbanksWiggins05c,SchubertWaalkensWiggins06,WaalkensSchubertWiggins08,
Komatsuzaki00,Komatsuzaki02,Komatsuzaki05}

\subsection{Phase space dividing surfaces:  definition and properties}

We are concerned with the rate of unimolecular reactions, either dissociation
or isomerization, at fixed  energy, for a 
system described by a time-independent,  $n$ degree-of-freedom (DOF) classical Hamiltonian.  
Points in the $2n$-dimensional system phase space $\calM  = \mathbb{R}^{2n}$ are denoted $\bx \in \calM$.
The system Hamiltonian is $H(\bx)$, and
the $(2n-1)$ dimensional  energy shell at energy $E$, $H(\bx) = E$, is denoted $\Sigma_E \subset \calM$. 
The corresponding microcanonical phase
space density is $\delta(E - H(\bx))$, and  
the associated density of states
for the complete energy shell at energy $E$ is
\begin{equation}
\rho(E) = \Int{\bx}{\calM}{} \delta(E - H(\bx)).
\end{equation}

The first step in the analysis is to define the region of the energy surface corresponding to 
the reactant of interest. For complex systems there may be many such regions 
of interest \cite{Wales03}, but for the moment we focus on only one such region, and the rate 
at which trajectories leave that particular region.  We treat either unimolecular dissociations
or isomerizations in which molecules are removed from consideration immediately after
passing from the reactant region of phase space, and are not
therefore directly concerned here with reactive flux correlation functions or
associated relaxation kinetics \cite{Dumont89,Dumont89a,Chandler78,Chandler87,Gray87}.

A standard approach to defining the reactant region 
is to consider the potential energy function--a configuration space based approach \cite{Wales03}.
In the simplest case the reactant region is
associated with a single local minimum, and trajectories exit/enter
the reactant region by passing over energetically accessible 
saddle points of the potential.
Although there could be several saddles controlling rates of exit  
from the potential well, as mentioned we will only consider the case of a single saddle. 
We comment on the case of multiple saddles below.

However, dynamics occurs in phase space (i.e., on the energy surface $\Sigma_E \subset \calM$
for the situation we are considering) while the picture described above is concerned 
with the projection of trajectories from phase space to configuration space. 
For multidimensional systems such as polyatomic molecules ($N \geq 3$ DoF), it is 
in general no longer possible to define or compute a dividing surface with desirable dynamical attributes
such as the no-recrossing property by
working in configuration space alone, and a phase space perspective is necessary
\cite{wwju,ujpyw,WaalkensBurbanksWiggins04,WaalkensWiggins04,WaalkensBurbanksWigginsb04,
WaalkensBurbanksWiggins05,WaalkensBurbanksWiggins05c,SchubertWaalkensWiggins06,WaalkensSchubertWiggins08,
Komatsuzaki00,Komatsuzaki02,Komatsuzaki05}.

While the phase space approach to reaction dynamics reviewed briefly below 
does not give a complete solution to the problem (questions of transition state bifurcations and 
nonintegrability remain to be addressed, for example),
it does provide an important component, 
which, when coupled with knowledge of certain properties of the system Hamiltonian function, 
does provide the necessary information to identify the reactant region in 
some situations of interest.  Saddle points of the potential energy surface 
still play a role -- in particular, 
rank one saddles. For Hamiltonian functions that are the sum of the kinetic energy 
and the potential energy a rank one saddle point of the potential energy function 
is manifested as an equilibrium point of saddle type for 
the $2n$ dimensional Hamilton's equations \cite{Wiggins94}. 
More precisely, it is an equilibrium point of saddle-center-$\ldots$-center stability type, 
meaning that the $2n \times 2n$ matrix associated with the linearization of 
Hamilton's equations about this equilibrium point has $2n$ eigenvalues of the form 
$\pm \lambda$, $\pm i \omega_k$, $k=2, \ldots, n$ ($\lambda, \, \omega_k >0$).  
The significance of saddle points of this type for Hamilton's equations  
is that, for a range of energies above that of the saddle 
(we explicitly discuss the range of energies later on), 
the energy surfaces have the {\em bottleneck property} in a phase space
neighborhood near the saddle, i.e. the $2n-1$ dimensional  
energy surface {\em locally} has the geometrical structure of the product 
of a $2n-2$ dimensional sphere and an interval, $S^{2n-2} \times I$. 
The bottleneck  property is significant because in the vicinity of the bottleneck 
we are able to construct a dividing surface, $\text{DS}(E)$,  
(where `$E$' denotes the energy dependence) with very desirable properties.  
For each energy in this range above the saddle $\text{DS}(E)$  
locally ``disconnects'' the energy surface into two, disjoint pieces 
with the consequence that the only way to pass from one 
piece of the energy surface to the other is to cross $\text{DS}(E)$. 
The dividing surface has the geometrical structure of a 
$2n-2$ dimensional sphere, $S^{2n-2}$, which is divided into two $2n-2$ dimensional hemispheres, 
denoted $\text{DS}_{\text{in}}(E)$ and $\text{DS}_{\text{out}}(E)$ that are joined at an equator,
which is a $2n-3$ dimensional sphere, $S^{2n-3}$. The hemisphere
$\text{DS}_{\text{in}}(E)$ corresponds to initial conditions of
trajectories that enter the reaction region while $\text{DS}_{\text{out}}(E)$ corresponds to 
initial conditions of trajectories that exit the reaction region, 
both by passing through the bottleneck in the energy surface. 
The equator $S^{2n-3}$ is an invariant manifold of saddle stability type, 
a so-called  {\em normally hyperbolic invariant manifold} (NHIM) \cite{Wiggins94}. 
The NHIM acts as the ``anchor'' for this entire construction and 
is of great physical significance: it is the actual ``saddle'' in phase space identified 
as the ``activated complex'' of reaction rate dynamics \cite{Pollak78,Truhlar96,WaalkensSchubertWiggins08}). 
Our focus here is on microcanonical rates,  and it 
has been shown that $\text{DS}_{\text{in}}(E)$ and $\text{DS}_{\text{out}}(E)$ 
have the essential \emph{no-recrossing property} and that the flux across them 
is minimal \cite{WaalkensWiggins04}. We denote the directional flux 
across these hemispheres by $\phi_{\text{in}} (E)$ and $\phi_{\text{out}} (E)$, 
respectively, and note that $\phi_{\text{in}} (E)+ \phi_{\text{out}}(E)=0$. 
For our purposes we only need the magnitude of the flux, and so set
$\vert \phi_{\text{in}} (E)\vert =  \vert \phi_{\text{out}}(E)\vert \equiv \phi (E)$.  
Most significantly,  the hemisphere  $\text{DS}_{\text{in}}(E)$ is the correct surface 
across which to compute the ``exact'' flux into the reaction region.

\subsection{Phase space volumes and gap times}
\label{subsec:volumes}

The disjoint regions of phase space corresponding to species A (reactant)
and B (product) will be denoted
$\calM_{\text{A}}$ and $\calM_{\text{B}}$, respectively \cite{footnote0}.
We assume that all coordinates and momenta are bounded on the 
reactant energy shell $\Sigma_E \cap \calM_{\text{A}}$, and that
it is possible to define a boundary (dividing surface) DS in phase space separating species A and B.  
As discussed above, the DS can be rigorously defined to be locally a surface of
no return (transition state).
The microcanonical density of states for reactant species A is
\begin{equation}
\rho_{\text{A}}(E) = \Int{\bx}{\calM_{\text{A}}}{} \delta(E - H(\bx)) 
\end{equation}
with a corresponding expression for the density of states $\rho_{\text{B}}(E)$
for product B for the case of compact product energy shell $\calM_{\text{B}}$.

Provided that the flow is everywhere transverse to $\text{DS}_{\text{in, out}}(E)$,
those phase points  in the reactant region $\calM_{\text{A}}$ 
that lie on crossing trajectories \cite{DeLeon81,Berne82} (i.e., that will react, and so are
``interesting'' in Slater's terminology \cite{Slater59})
can be specified uniquely by coordinates $(\bqb, \bpb, \psi)$, 
where $(\bqb, \bpb) \in \text{DS}_{\text{in}}(E)$ is a point on
$\text{DS}_{\text{in}}(E)$, the incoming half of the DS, specified by $2(n-1)$ 
coordinates $(\bqb, \bpb)$, and
$\psi$ is a time variable.  (Dividing surfaces constructed by
the normal form algorithm are guaranteed to be transverse to the vector field, except at the NHIM, where 
the vector field is tangent \cite{wwju,ujpyw}.)
The point $\bx(\bqb, \bpb, \psi)$ is reached by propagating the
initial condition $(\bqb, \bpb) \in \text{DS}_{\text{in}}(E)$ forward for time $\psi$
(see Figure \ref{phase_space_1}).
As all initial conditions on $\text{DS}_{\text{in}}(E)$
(apart from a set of trajectories of measure zero lying on stable manifolds)
will leave the reactant region in finite time by crossing $\text{DS}_{\text{out}}(E)$, for each 
$(\bqb, \bpb) \in \text{DS}_{\text{in}}(E)$ we can define the \emph{gap time} 
$s = s(\bqb, \bpb)$, which is the
time it takes for the incoming trajectory to traverse the reactant region.
That is, $\bx(\bqb, \bpb, \psi = s(\bqb, \bpb)) \in \text{DS}_{\text{out}}(E)$.
For the phase point $\bx(\bqb, \bpb, \psi)$, we therefore have
$0 \leq \psi \leq s(\bqb, \bpb)$.

The coordinate transformation $\bx \to (E, \psi, \bqb, \bpb)$ is canonical
\cite{Arnold78,Thiele62,Binney85,Meyer86}, so that the phase space volume element is
\begin{equation}
\label{coord_1}
\rmd^{2n} \bx = \rmd E \, \rmd \psi  \, \rmd \sigma  
\end{equation}
with $\rmd \sigma \equiv \rmd^{n-1} \qb \, \rmd^{n-1} \pb$ 
an element of $2n-2$ dimensional area on the DS.

As defined above, the magnitude $\phi(E)$ of the flux through dividing surface 
$\text{DS}(E)$ at energy $E$ is given by 
\begin{equation}
\label{flux_1}
\phi(E) = \left\vert\Int{\sigma}{\text{DS}_{\text{in}}(E)}{} \right\vert,
\end{equation}
where the element of area $\rmd \sigma$ is precisely the restriction to DS of the
appropriate flux $(2n-2)$-form $\omega^{(n-1)}/(n-1)!$ corresponding to the Hamiltonian vector field 
associated with $H(\bx)$ \cite{Toller85,Mackay90,Gillilan90,WaalkensWiggins04}.
The reactant phase space volume occupied by points initiated on the dividing surface
$\text{DS}_{\text{in}}$ with energies between $E$ and $E + \rmd E$ is therefore
\cite{Thiele62,Brumer80,Pollak81,Binney85,Meyer86,WaalkensBurbanksWiggins05,WaalkensBurbanksWiggins05c} 
\begin{subequations}
\label{vol_1}
\begin{align}
\rmd E \Int{\sigma}{\text{DS}_{\text{in}}(E)}{} \Int{\psi}{0}{s} 
& = \rmd E \Int{\sigma}{\text{DS}_{\text{in}}(E)}{}  s \\
&= \rmd E \, \phi(E) \, \sbar 
\end{align}
\end{subequations}where the \emph{mean gap time} $\sbar$ is defined as
\begin{equation}
\sbar = \frac{1}{\phi(E)} \, \Int{\sigma}{\text{DS}_{\text{in}}(E)}{}  s
\end{equation}
and is a function of energy $E$.
The reactant density of states $\rho^{\text{C}}_{\text{A}}(E)$
associated with crossing trajectories only (those trajectories that enter and exit
the reactant region \cite{Berne82}; see below) is then
\begin{equation}
\label{vol_1p}
\rho^{\text{C}}_{\text{A}}(E) = \phi(E) \, \sbar
\end{equation}
where the superscript $\text{C}$ indicates the restriction to crossing trajectories. 
The result \eqref{vol_1p} is essentially the content of the so-called 
classical spectral theorem 
\cite{Brumer80,Pollak81,Binney85,Meyer86,WaalkensBurbanksWiggins05,WaalkensBurbanksWiggins05c}.

If \emph{all} points in the reactant region of phase space eventually react (that is, 
all points lie on crossing trajectories \cite{DeLeon81,Berne82}) then
$\rho^{\text{C}}_{\text{A}}(E) = \rho_{\text{A}}(E)$, the full reactant phase space density of states.
Apart from a set of measure zero, all phase points $\bx \in \calM_{\text{A}}$ 
can be classified as either trapped (T) or crossing (C) \cite{Berne82}.
(Further discussion of this division of the reactant phase space in terms of 
the Poincar\'{e} recurrence theorem is given in Appendix \ref{app:recurrence}.)
A phase point in the trapped region $\calM_{\text{A}}^{\text{T}}$ never crosses the DS, 
so that the associated trajectory does not contribute to the reactive flux.
Phase points in the crossing region $\calM_{\text{A}}^{\text{C}}$ do however eventually
cross the dividing surface, and so lie on trajectories that contribute to the reactive flux.
In general, however, as a consequence of the existence of trapped trajectories
(either trajectories on invariant \emph{trapped} $n$-tori \cite{DeLeon81,Berne82} or 
trajectories asymptotic to other invariant objects of zero measure), 
we have the inequality \cite{Thiele62,Berne82} 
\begin{equation}
\label{vol_2}
\rho_{\text{A}}^{\text{C}}(E) \leq \rho_{\text{A}}(E).
\end{equation}
If $\rho_{\text{A}}^{\text{C}}(E) < \rho_{\text{A}}(E)$, then it is in principle necessary to
introduce corrections to statistical estimates of reaction rates 
\cite{Berne82,Gray87,Berblinger94,Stember07}.
Numerical results for $\rho^{\text{C}}(E)$ and $\rho(E)$ for the HCN molecule are discussed 
below \cite{WaalkensBurbanksWigginsb04,WaalkensBurbanksWiggins05}.

\subsection{Gap time and reactant lifetime distributions}
\label{subsec:gaps}

Of central interest is the \emph{gap time distribution}, $\calP(s; E)$: the probability
that a phase point on $\text{DS}_{\text{in}}(E)$ at energy $E$ has a gap time between 
$s$ and $s +\rmd s$ is equal to $\calP(s; E) \rmd s$. 
An important idealized gap distribution is the random, exponential distribution
\begin{equation}
\label{exp_1}
\calP(s; E) = \Bbbk(E) \, e^{-\Bbbk(E) s}
\end{equation}
characterized by a single decay constant $\Bbbk$ (where $\Bbbk$ depends on energy $E$), 
with corresponding mean gap time $\sbar = \Bbbk^{-1}$.

The lifetime (time to cross the dividing surface $\text{DS}_{\text{out}}(E)$) 
of phase point $\bx(\bqb, \bpb, \psi)$ is
$t = s(\bqb, \bpb) - \psi$ (cf.\ Fig.\ \ref{phase_space_1}b).  
The volume of reactant phase space occupied by
trajectories having lifetimes $t \geq t'$ at energy $E$ is then
\begin{equation}
\text{Vol}(t \geq t'; E) = \phi(E) \, \Int{s}{t'}{+\infty}  (s - t')\, \calP(s; E)
\end{equation}
so that the corresponding probability of an interesting phase point in the reactant region having
a lifetime $t \geq t'$ is obtained by dividing this volume by the total volume
occupied by points on crossing trajectories, $\phi(E) \sbar$,
\begin{equation}
\text{Prob}(t \geq t'; E) 
= \frac{1}{\sbar} \, \Int{s}{t'}{+\infty} \calP(s; E) (s - t').
\end{equation}
The corresponding reactant \emph{lifetime distribution function} $\bbP(t; E)$ at energy $E$
is therefore 
\begin{subequations}
\label{life_1}
\begin{align}
\bbP(t; E) &= -\frac{\rmd}{\rmd t'}\; \text{Prob}(t \geq t'; E) \Big\vert_{t'=t} \\
&= \frac{1}{\sbar} \, \Int{s}{t}{+\infty} \calP(s; E)
\end{align}
\end{subequations}
where the fraction of interesting (reactive) phase points having lifetimes between $t$ and $t + \rmd t$ is
$\bbP(t; E) \rmd t$.  It is straightforward to verify that 
the lifetime distribution \eqref{life_1} is normalized.
Note that an exponential gap distribution \eqref{exp_1} 
implies that the reactant lifetime
distribution $\bbP(t; E)$ is also exponential; both gap and lifetime distributions 
for realistic molecular potentials have 
been of great interest since the earliest days of trajectory simulations of 
unimolecular decay \cite{Bunker66,Bunker73}.

We emphasize that the rigorous relation \eqref{life_1} 
between the gap time distribution and the reactant lifetime distribution follows quite 
straightforwardly from 
our continuous time formulation of the problem using the canonical transformation of
phase space variables eq.\ \eqref{coord_1} and the properties of the dividing
surfaces $\text{DS}$ (cf.\ ref.\ \onlinecite{Dumont86}).   

A concise derivation of the delay differential equation for the Delayed Lifetime Gap Model
\cite{Dumont86} is presented in Appendix \ref{app:DLGM}.

\subsection{Reaction rates}

We start with the (classical) expression 
for the rate $k(T)$ of a collisionally activated unimolecular process at temperature $T$ 
derived by Thiele \cite{Slater56,Slater59,Thiele62}.  Using the notation established above, the 
rate coefficient $k(T)$ is 
\begin{subequations}
\label{k1}
\begin{align}
k(T) &= \frac{1}{Z_{\text{A}}} \Int{E}{E_0}{+\infty} e^{-\beta E} 
\Int{\sigma}{\text{DS}_{\text{in}}(E)}{} \left[ 1 - e^{-\omega s}\right] \\
\label{k1b}
& = \frac{1}{Z_{\text{A}}} \Int{E}{E_0}{+\infty} e^{-\beta E} 
\, \phi(E) \left[ 1 - \overline{e^{-\omega s}} \right].
\end{align}
\end{subequations}
Here, 
$Z_{\text{A}}(T)$ is the reactant partition function
\begin{equation}
Z_{\text{A}} = \Int{E}{}{} e^{-\beta E} \rho_{\text{A}}(E),
\end{equation}
$\beta = 1/k_{\text{B}}T$, $\omega$ is the effective collision rate per molecule, 
$E_0$ is the threshold energy for reaction, and the overline in eq.\ \eqref{k1b}
denotes an average
over the dividing surface $\text{DS}_{\text{in}}(E)$.  The physical interpretation of
expression \eqref{k1} is that the thermal  reaction rate $k(T)$ is given by the
average of the equilibrium activation rate times the probability that an 
activated phase point will react (that is, cross the dividing surface DS$_{\text{out}}$)
before it suffers a (strong) deactivating collision.

We note that the rate expression \eqref{k1} makes sense even though
for larger energies $E$ the dividing surface DS$_{\text{in}}(E)$ 
with desired dynamical properties might no longer exist, 
as the contribution from high energies is damped away by the exponential Boltzmann factor.
(Thiele \cite{Thiele62} assumed the existence of a suitable dividing surface for all
energies $E$.)

The limiting expressions for $k(T)$ obtained at high and low pressures are of interest.
At high pressures ($\omega \to \infty$) we have
\begin{subequations}
\begin{align}
k_\infty(T) \equiv \lim_{\omega\to \infty}k(T) 
& = \frac{1}{Z_{\text{A}}} \Int{E}{E_0}{+\infty} e^{-\beta E}\,
\Int{\sigma}{\text{DS}_{\text{in}}(E)}{}  \\
& = \frac{1}{Z_{\text{A}}} \Int{E}{E_0}{+\infty} e^{-\beta E} \, \phi(E)  \\
& = \frac{1}{Z_{\text{A}}} \Int{E}{E_0}{+\infty} \rho_{\text{A}}(E) e^{-\beta E} 
\, k^{\text{RRKM}}_f(E)
\end{align}
\end{subequations}
where the quantity
\begin{equation}
\label{k_RRKM}
k^{\text{RRKM}}_f(E) \equiv \frac{\phi(E)}{\rho_{\text{A}}(E)}
\end{equation}
is the statistical (RRKM) microcanonical rate for the forward reaction
(A $\to$ B) at energy $E$, the ratio of the magnitude of
the flux $\phi(E)$ through $\text{DS}_{\text{in}}(E)$ 
to the total reactant density of states \cite{Robinson72,Forst73}.
The rate $k_\infty(T)$ is then the canonical average of the microcanonical statistical
rate $k^{\text{RRKM}}_f(E)$.
(The collision rate $\omega$ should not be so large that
trajectories of systems crossing the dividing surface are significantly perturbed by collisions.)

Clearly, if $\rho_{\text{A}}(E) = \rho_{\text{A}}^{\text{C}}(E)$, then
\begin{equation}
k^{\text{RRKM}}_f(E) = \frac{1}{\sbar}
\end{equation}
the inverse mean gap time.
In general, the inverse of the mean gap time is
\begin{subequations}
\label{k2}
\begin{align}
\frac{1}{\sbar} &= \frac{\phi(E)}{\rho_{\text{A}}^{\text{C}}} \equiv k^{\text{RRKM}}_{f, \text{C}} \\
& = k^{\text{RRKM}}_f \, \left[\frac{\rho_{\text{A}}(E)}{\rho_{\text{A}}^{\text{C}}(E)}\right] \\
& \geq k^{\text{RRKM}}_f.
\end{align}
\end{subequations}
The rate $k^{\text{RRKM}}_{f, \text{C}}$ can be interpreted as the 
statistical unimolecular reaction  rate corrected for the volume of trapped trajectories in the
reactant phase space \cite{Dumont86,Berne82,Gray87,Berblinger94}.  
The modified statistical rate is therefore predicted to be \emph{greater} than the RRKM rate,
a prediction usually at odds with numerical simulations (cf.\ results presented 
below, also refs \onlinecite{Berne82,Gray87,Berblinger94,Carpenter05}).

The low pressure ($\omega \to 0$) limit of the rate is
\begin{subequations}
\begin{align}
k_0(T) \equiv \lim_{\omega\to 0}k(T) & = \frac{\omega}{Z_{\text{A}}} \Int{E}{E_0}{+\infty} e^{-\beta E}\,
\Int{\sigma}{\text{DS}_{\text{in}}(E)}{} s \\
& = \frac{\omega}{Z_{\text{A}}} \Int{E}{E_0}{+\infty} e^{-\beta E} \, \phi(E) \sbar \\
& = \frac{1}{Z_{\text{A}}} \Int{E}{E_0}{+\infty} \rho_{\text{A}}(E) e^{-\beta E} 
\, k^{\text{RRKM}}_f(E) \, \omega \sbar
\end{align}
\end{subequations}
showing that the effective microcanonical rate is smaller than
the RRKM statistical weight by a factor $\omega \sbar \ll 1$.
That is, in the low pressure limit the reaction rate is proportional to the rate 
of collisional activation $\omega$, with each molecule taking a time 
$\sbar$ on average to react.

Different choices of transition state location will in general 
result in different mean gap times \cite{Dumont86}.
We emphasize that exact and unambiguous calculation of  
the mean gap time (\ref{k2}) is possible given knowledge of
the {\em phase space} 
geometrical structures that enable us to construct the reaction region, a dividing 
surface with minimal flux (hence the exact flux can be computed without integrating trajectories), 
and the reactive volume, i.e., the volume of the energy surface corresponding to 
interesting initial conditions $\bx \in \calM_{\text{A}}^{\text{C}}$.  The fundamental geometrical structures 
required to compute these quantities are the $2n-2$ dimensional
hemipheres $\text{DS}_{\text{in}}(E)$ and $\text{DS}_{\text{out}}(E)$ that control 
the entrance to the reaction region and exit from the reaction region, respectively. 
These geometrical structures are what we use to compute flux 
and we sample initial conditions on $\text{DS}_{\text{in}}(E)$ and integrate them 
until they reach $\text{DS}_{\text{out}}(E)$ in order to compute $\sbar_{\text{DS}_{\text{in}}(E)}$. 
A detailed algorithm has previously been given for computing the 
dividing surfaces and the flux across them 
\cite{WaalkensBurbanksWiggins04,WaalkensWiggins04,WaalkensSchubertWiggins08}.
In these references, numerical tests were also described 
for determining the range of energies above the 
saddle for which the dividing surface ``locally disconnects'' the energy surface 
in the way described above (in particular, for energies sufficiently larger than 
the saddle the dynamics may not ``feel'' the influence of the saddle point at all and 
the energy surface could deform so much that it would make no sense to speak of 
disjoint regions of ``reactants'' and ``products'' \cite{Li06}).

\subsection{Multiple saddles}

So far, we have only considered a reaction region where access in and out of the 
region is controlled by a single saddle. We can similarly consider a reaction region  
(in phase space) where access in and out of the region is controlled by $d$ saddles 
(actually, $d$ saddle-center- $\ldots$ - center type equilibria). 
Associated with each saddle we can compute a $2n-2$ dimensional dividing surface, $\rm \text{DS}_i (E)$, 
that is divided into two $2n-2$ dimensional hemispheres 
$\text{DS}_{\text i, \text{in}}(E)$ and $\text{DS}_{\text i,  \text{out}}(E)$  
which have the same interpretation as above. The magnitude of the flux out 
of the reactive region is denoted by $\sum_{i=1}^{d} \phi_{\text i}(E)$ 
and it is shown in \cite{WaalkensBurbanksWiggins05,WaalkensBurbanksWiggins05c}
that the volume of initial conditions in the energy surface that correspond to
trajectories that leave the reaction region is given by 
$\sum_{i=1}^{d} \sbar_{\text{DS}_{\text i, \text{in}}(E)} \,\phi_{\text i}(E)$, 
where $ \sbar_{\text{DS}_{\text i, \text{in}}(E)}$ is the average time 
for trajectories starting on $\text{DS}_{\text i, \text{in}}(E)$ to cross 
$\text{DS}_{\text j, \text{out}}(E)$, 
for any $1 \le j \le d$. In this case the corrected total statistical escape rate is given by:
\begin{equation}
k^{\text{RRKM}}_{f, \text{C}}(E; d) = \frac{\sum_{i=1}^{d} \phi_{\text i}(E)}{\sum_{i=1}^{d} 
\sbar_{\text{DS}_{\text i, \text{in}}(E)} \,\phi_{\text i}(E)}.
\label{rate_d}
\end{equation}

An important case is that where all of the saddles are symmetric in the sense 
that $\phi_{\text i}(E)=\phi_{\text j}(E)$, 
$\sbar_{\text{DS}_{\text i, \text{in}}(E)} \,\phi_{\text i, \text{in}}(E)
= \sbar_{\text{DS}_{\text j, \text{in}}(E)} \,\phi_{\text j, \text{in}}(E)$, 
for all $1 \le i, j \le d$ then the corrected statistical rate 
\eqref{rate_d} reduces to expression \eqref{k2}. 

The symmetric situation applies to the case of HCN isomerization considered in
\cite{WaalkensBurbanksWigginsb04,WaalkensBurbanksWiggins05}.
Numerical results for HCN are further discussed in Sec.\ \ref{sec:HCN_data} of the present paper.

\newpage
\section{HCN isomerization dynamics}
\label{sec:HCN_data}

The isomerization dynamics of HCN, HCN $\rightleftharpoons$ CNH, 
has been widely studied, using both classical and quantum mechanics: see, for example, refs 
\onlinecite{Gray80,pmpb1,pmpb2,Lan93,Bentley93,Tang94,Shah99,WaalkensBurbanksWigginsb04,WaalkensBurbanksWiggins05,Li05,Gong05a}
and references therein.
In the calculations reported in
refs \onlinecite{WaalkensBurbanksWigginsb04,WaalkensBurbanksWiggins05},
the potential energy surface of Murrell, Carter, and Halonen (MCH) \cite{MCH} was used. 
For the MCH potential energy surface the saddle point is at energy -12.08 eV and the 
trajectory calculations 
in refs \onlinecite{WaalkensBurbanksWigginsb04, WaalkensBurbanksWiggins05} were all carried out at energy
0.2 eV above the saddle. 

The HCN molecule is modelled as a planar system with zero angular momentum, so that there are 
three DoF \cite{WaalkensBurbanksWigginsb04,WaalkensBurbanksWiggins05}. 
In planar HCN there are two saddles, related by reflection 
symmetry, separating reactant (HCN) from product (CNH), 
with bond angle $\gamma = \pm \gamma^\ast \simeq \pm 67^\circ$ 
(see Figure \ref{hcn_pes}).
The mean gap time is found to be $\sbar = 0.174 \, {\rm ps}$ which 
corresponds to an isomerization rate of $0.14 \times 10^{-3}$ a.u. 
\cite{WaalkensBurbanksWigginsb04,WaalkensBurbanksWiggins05} (see below).
A discussion of numerical aspects and efficiency of the calculations 
was also given in refs \onlinecite{WaalkensBurbanksWigginsb04,WaalkensBurbanksWiggins05}.

\subsection{Gap time and reactant lifetime distributions}

The phase space structures of interest at fixed energy $E$, namely the NHIMs 
and the dividing surfaces separating reactant from product, are 
computed via an algorithmic procedure based on Poincar{\'e}-Birkhoff normalization that is
described in refs \onlinecite{ujpyw, WaalkensSchubertWiggins08}. 
The Poincar{\'e}-Birkhoff normalization
provides a nonlinear, symplectic transformation from the 
original, physical coordinates $(q, p)$ to a
new set of coordinates, the \emph{normal form coordinates} $(\qb, \pb)$,
in terms of which the dynamics is ``simple''.
Morover, in these normal form coordinates the phase space structures governing reaction dynamics 
can be expressed in terms of explicit
formulae, as described in refs \onlinecite{ujpyw, WaalkensSchubertWiggins08}.
Their influence on the dynamics, in the normal form coordinates,
is very easy to understand, and the geometrical structures can then be
mapped back into the original coordinate system.

The Poincar\'e-Birkhoff normal form theory
provides an algorithm to compute the symplectic tranformation $T$ from physical 
coordinate to normal coordinates,
\begin{equation}
T(q, p) = (\qb, \pb).
\end{equation}
In a local neighbourhood $\mathcal{L}$ of the equilibrium point of interest, 
this trandformation ``unfolds'' the dynamics into a ``reaction coordinate'' and ``bath
modes'': expressing the system Hamiltonian $H$ in the new coordinates, $(\qb, \pb)$, via
\begin{equation}
H_\text{NF}(\qb, \pb) = H(T^{-1} (q, p)),
\end{equation}
gives $H_\text{NF}$ in a simplified form. This
``unfolding'' into a reaction coordinate and bath modes is one way
of understanding how we are able to construct the phase space
structures, in the normal form coordinates, that govern the
dynamics of reaction.  The explicit
expressions for the coordinate transformations, $T(q, p) = (\qb, \pb)$
and $T^{-1}(\qb, \pb) = (q, p)$, between the normal form (NF)
coordinates and the original coordinates provided by the normalization procedure are
also essential, as they allow  us to 
transform the phase space structures constructed in 
normal form coordinates back into the original physical coordinates.

We consider an ensemble of trajectories with initial conditions sampled uniformly 
on one of the two dividing surfaces,  $\text{DS}_{1, \text{in}}(E)$ say, 
according to
the measure $\rmd \sigma$ (cf.\ eq.\ \eqref{flux_1}), so that all trajectories 
initially enter the reactant region of phase space via channel 1.  
Initial conditions on the dividing surface are 
obtained by uniformly sampling the $2n - 2$ dimensional DS 
in normal form coordinates ($\bqb, \bpb$) with reaction coordinate variables $(\qb_1, \pb_1)$
determined by energy conservation ($E = -11.88$ eV) 
together with the constraint $\qb_1 = \pb_1$.
The inverse of the normal form coordinate transformation is used to compute 
physical coordinates for initial phase points, and trajectories are propagated forward in time
in physical coordinates.  As a trajectory approaches 
either dividing surface $\text{DS}_{i, \text{out}}(E)$,
a transformation is made to normal form coordinates, which allows accurate determination
of the time at which the trajectory crosses the out DS.  This is the gap time.

The ensemble consists
of 815871 trajectories; of these, 83599 ultimately exit through channel (DS) 1, while 
732272 exit through channel 2.  The branching ratio 
for the whole ensemble is then $8.76$, very different from the 
statistical value unity dictated by symmetry.  
The mean gap time for the complete
ensemble is 0.174 ps \cite{WaalkensBurbanksWiggins05}.
For trajectories 
reacting via channel 1, $\sbar = 0.714$ ps, while $\sbar = 0.112$ ps for those reacting via channel 2.
Those trajectories that exit via the same transition state through which they entered therefore
have a significantly larger average gap time.  This makes physical sense, as such trajectories 
must have at least one turning point in the bending motion.

The gap time distribution $\calP(s; E)$ 
for the complete ensemble at constant $E = -11.88$ eV is shown in Figure \ref{gap_fig}a
(cf.\ Figure 4b of ref.\ \onlinecite{WaalkensBurbanksWiggins05}).  
Gap time distributions for subensembles 1 and 2 are shown in Figures
\ref{gap_fig}b and \ref{gap_fig}c, respectively.
In addition to the gap time distribution itself, we also consider the 
cumulative distribution $F(t)$, which is defined as
the fraction of trajectories on the DS with gap times $s \geq t$, and is 
simply the product of the normalized reactant lifetime distribution function $\bbP(t; E)$ 
and the mean gap time $\sbar$ (cf.\ eq.\ \eqref{life_1}):
\begin{subequations}
\begin{align}
F(t) & = \Int{s}{t}{+\infty} \calP(s; E) \\
& = \sbar \, \bbP(t; E).
\end{align}
\end{subequations}
For the random gap time distribution \eqref{exp_1}, the cumulative gap time
distribution is exponential, $F(t) = e^{-\Bbbk t}$.

Cumulative gap time distributions for $0 \leq t \leq 0.4$ ps are shown 
for the whole ensemble in Fig.\ \ref{short_life_fig}a, and for the two subensembles 
in Figs \ref{short_life_fig}b and \ref{short_life_fig}c, respectively.
Reactant lifetime distributions for the whole ensemble 
over longer time intervals ($0 \leq t \leq 25$ ps)
are presented in Figure \ref{lifetime_all}; both $\log[F(t)]$ vs $t$ and $\log[F(t)]$ vs $\log t$ 
plots are shown.  Corresponding plots for the two subensembles are shown in
Figs \ref{lifetime_1} and \ref{lifetime_2}.

We now discuss properties of the gap time and lifetime distributions on 
various physically relevant timescales:

\begin{enumerate}
\item Very short times ($t \ll \sbar$)

By construction, the phase space DS used to separate reactants and products 
eliminates local (short-time) recrossings 
\cite{wwju,ujpyw,WaalkensBurbanksWiggins04,WaalkensWiggins04,WaalkensBurbanksWigginsb04,
WaalkensBurbanksWiggins05,WaalkensBurbanksWiggins05c,SchubertWaalkensWiggins06,WaalkensSchubertWiggins08}.
Hence, there are no very short gaps.

\item  Short times ($t \simeq \sbar$).

The gap time distribution for the complete ensemble shows ``pulses'' of reacting
trajectories.  Each pulse is associated with a bundle of trajectories
that execute a certain number of oscillations in the reactant well
before crossing one or the other DS \cite{WaalkensBurbanksWigginsb04,WaalkensBurbanksWiggins05}.
The first pulse
is associated with trajectories exiting via channel 2, the second pulse with trajectories
exiting via channel 1, and so on.  Similar structure has been seen in the lifetime distribution
computed for escape of Rydberg electrons in crossed fields \cite{Gabern05}.

The cumulative gap time distributions up to times $\simeq \sbar$ 
(Fig.\ \ref{short_life_fig}) exhibit a structured and
faster-than-exponential
decay and so cannot readily be fitted to an exponential curve in order to
obtain an effective decay rate for trajectories leaving
the HCN well.

\item  Intermediate times $10 \sbar \gtrsim t \gtrsim \sbar$.

On intermediate timescales, the peaks associated in the gap time
distribution associated with individual pulses begin to overlap.
The reactant lifetime distribution exhibits \emph{algebraic} (power law) decay,
$F(t) \sim 1/t^\alpha$, with $\alpha \simeq 0.705$ (Figure \ref{lifetime_all}a).
Such power law decay has been seen in other models for isomerization 
\cite{Shojiguchi07,Shojiguchi07a}, and is in general associated 
with fractional kinetics \cite{Zaslavsky05}.

Attempts have been made to rationalize the existence of power law lifetime distributions 
in such systems in terms of a hierarchical set of bottlenecks, perhaps associated with 
the Arnold web \cite{Lichtenberg92,Shojiguchi08} presumed to exist in the vicinity of
the minimum of the potential well.  At this stage, however, a more quantitative explanation 
of the dynamical origins of algebraic decays such as those seen here at intermediate times
remains an open problem.

On the same timescale, both subensembles exhibit algebraic decay with essentially
\emph{identical} exponents $\alpha$:  $0.708$ and $0.701$, respectively 
(Figs \ref{lifetime_1}a, \ref{lifetime_2}a).  
Of course, if one starts with an ensemble of reactant phase points whose distribution
possesses the symmetry of the phase space induced by the reflection symmetry of
the potential, then equality of the exponents for the algebraic portion of the decay
is expected.   Our ensemble of initial conditions is nevertheless
highly asymmetric.  We comment on the observed exponent equality further below 
in our discussion of branching ratios.

\item  Long times $t \gg \sbar$.

At longer times the lifetime distribution exhibits exponential decay, $F(t) \sim e^{-k t}$, with
exponent $k \simeq 0.092$ ps$^{-1}$ (the decay constant is obtained by fitting 
the data for $10 \leq t \leq 20$ ps, cf.\ Figure \ref{lifetime_all}b).
Decay constants $k$ are found to be 
identical for trajectories exiting through 
either channel ($0.093$ ps$^{-1}$  for channel 1, Fig \ref{lifetime_1}b, $0.091$ ps$^{-1}$ 
for channel 2, Fig \ref{lifetime_2}b).  Again, equality of decay rates is
expected on symmetry grounds for a symmetric ensemble.
More generally, the equality of decay rates 
for trajectory subensembles reacting via distinct channels is an implicit assumption
of statistical theories \cite{Dumont86}.  
An informative discussion of this equality is given in 
Sec.\ V(d) of ref.\ \onlinecite{Dumont86}.

For a system such as HCN with 2 identical transition states
the total forward decay constant $k = k_f$.

\end{enumerate}

\subsection{Statistical and modified statistical rates}

Having characterized the behavior of the gap time and lifetime distributions on various timescales, we now
consider the relation of the numerically determined decay  rate to various statistical estimates.

Figure \ref{survival} shows the survival probability $P_{\text{S}}(t)$ for phase points in the reactant 
region of phase space (cf.\ Fig.\ 3b of ref.\ \onlinecite{WaalkensBurbanksWiggins05}).
This quantity is computed by Monte Carlo sampling phase points $\bx$ in the reactant region,
$\bx \in \Sigma_E \cap \calM_{\text{A}}$,
and propagating trajectories until they either react through either channel or a cutoff time is reached.
The fraction of phase points surviving until time $t$ is $P_{\text{S}}(t)$.  It can be seen that
$P_{\text{S}}(t)$ appears to converge relatively slowly to a constant value 
$f_{\text{T}} = P_{\text{S}}(\infty) \simeq 0.91$ \cite{footnote3}.  This shows that
over $90$\% of phase points are trapped in the reactant region, and that the
density of states $\rho_{\text{A}}^{\text{C}} = (1- f_{\text{T}}) \rho_{\text{A}}$ associated
with crossing trajectories is approximately 10\% of the full reactant density of states $\rho_{\text{A}}$.

The value of $\rho_{\text{A}}^{\text{C}}$ has also been computed using the relation
\begin{equation}
\rho_{\text{A}}^{\text{C}} = 2 \, \sbar \phi(E),
\end{equation}
and shown to be identical with the result obtained via the survival 
probability \cite{WaalkensBurbanksWiggins05}.

Numerical values of relevant quantities are $\sbar = 0.163$ ps, 
$\phi = 0.00085 \times h^2$, where $h$ is Planck's constant,
$\rho_{\text{A}} = 0.795 \times h^3/\text{eV}$ and 
$\rho_{\text{A}}^{\text{C}} = 0.0715 \times h^3/\text{eV}$.
Computed values for the total statistical decay rate coefficient (in the symmetric case,
equal to the single channel rate constant $k_f$) are 
$k^{\text{RRKM}}_f =  0.517 \, \text{ps}^{-1}$ (RRKM, uncorrrected) and
$k^{\text{RRKM}}_{f, \text{C}} =  1/\sbar = 5.75 \, \text{ps}^{-1}$ (RRKM, corrected).
These values are to be compared with the long time decay rate $k = 0.092 \,\text{ps}^{-1}$ .

It is immediately apparent that, as noted previously 
(ref.\ \onlinecite{Berne82}, although cf.\ ref.\ \onlinecite{Gray87}), 
the value of the
statistical rate constant ``corrected'' for the volume of trapped reactant phase points
is both much larger than the uncorrected statistical rate and in significantly greater
disagreement with the exact (numerical) value of $k$.  
In fact, the ratio $k^{\text{RRKM}}_f/k = 5.62$, so that the uncorrected statistical
rate coefficient is within a factor of $6$ of the numerical
escape rate.  This is presumably due to compensating errors in the statistical 
calculation \cite{Berne82}. The presence of trapped regions of reactive phase space
decreases the volume of phase space that is available for reactive trajectories to explore; this
effect tends to increase the value of the actual escape rate with respect to the 
statistical estimate.  As can be seen from the numerical values given above, if this were the only
factor affecting the rate then the actual rate would be a factor of 10 larger than
the RRKM estimate.  However, additional dynamical trapping mechanisms, that are as yet
not fully understood for multimode systems \cite{Toda02,Shojiguchi08,Paskauskas08}, 
serve to delay the exit of phase points from the reactant region.  The competition between
these two effects then results in a value of the numerical escape rate that is
fairly close to the simple RRKM estimate \cite{Berne82}.

\subsection{Statistical branching ratio accompanied by nonexponential decay}

In Figure \ref{nlog12} we plot $\log [N(t)]$ versus $t$ for each subsensemble, 
where $N(t)$ is the number of phase points remaining in the well at time $t$.
Except at extremely short times $\lesssim 0.5$ ps, the
two curves are essentially identical.  For those times during which the 
decay curves are identical, the associated product branching ratio is
equal to its statistical value, unity.  The fact that the curves can be overlaid 
even at times for which the decay of each ensemble is nonexponential 
(algebraic) then implies that we have a statistical product ratio
in the absence of exponential decay.  (Note that exponential decay is
usually taken to be the signature of ``statistical'' dynamics \cite{Thiele62,Dumont86}.)
The identity of the decay curves also implies the equality 
of the algebraic decay exponents noted previously.

A qualitative explanation of this behavior is as follows:
note that we start with a highly asymmetric initial ensemble, 
entering via $\text{DS}_{1, \text{in}}$ only.
This bundle of trajectories passes through the HCN well, over to $\text{DS}_2$.  Some fraction
of the ensemble exits through $\text{DS}_{2, \text{out}}$ and is lost.  
The rest of the trajectories then turn back, and pass through the 
well again, over to $\text{DS}_1$.  Some fraction is again lost.  (These are the ``pulses''
seen in the short-time gap time distribution.) And so on.

The point is that, after the trajectory bundle has oscillated  back and forth in the HCN well several times, 
the set of phase points still in the well behaves \emph{as if} 
as if it were an ensemble consisting of trajectories initiated 
in equal numbers on both entry dividing surfaces and then propagated for $t \gtrsim 0.5$ ps.  
The underlying idea here is that, when trajectories are ``turned back'' from a DS, their subsequent time 
evolution can be qualitatively similar to that of 
trajectories actually initiated on the same DS.
For example, trajectories turned back at  $\text{DS}_2$ that are 
just ``outside'' the incoming reactive cylinder manifold associated with the
dividing surface \cite{DeAlmeida90,ujpyw,WaalkensSchubertWiggins08}
will track (shadow \cite{Guckenheimer83}) trajectories that actually enter 
the reactant region through $\text{DS}_{2, \text{in}}$ inside the reactive 
cylinder and lie close to the boundary (unstable manifold $W^{u}$).

The HCN gap time/lifetime distributions therefore imply that the initial
ensemble effectively attains the full symmetry of the reactant phase space some time before
it fully relaxes to the stage where exponential decay is observed \cite{Dumont86},
so that the system exhibits statistical branching ratios in the absence of 
exponential decay.

\subsection{Comparison with other calculations}

The trajectory calculations analyzed here were carried out using the MCH potential surface 
\cite{MCH} at a constant energy $E=-11.88$ eV ($0.2$ eV above the saddle energy).
Tang, Jang, Zhao and Rice (TJZR) 
\cite{Tang94} have carried out classical trajectory calculations using the same potential
surface for a number of energies, and have applied an approximate version \cite{Zhao92a} of
the 3-state statistical theory of Gray and Rice \cite{Gray87} as well as a reaction path
approach \cite{Jang93} to compute isomerization rate constants for the 3 DoF, zero angular
momentum HCN isomerization.

The energy value used by TJZR closest to that of the present work is $E = -11.5$ eV.
At this energy, TJZR extract an isomerization rate of $0.146 \times 10^{-3}$ a.u.\ from 
their trajectory calculations, corresponding to 
a mean lifetime of $0.166$ ps  \cite{Tang94}.
This rate, which determines the timescale for decay of the CNH population and is therefore strictly 
the relaxation rate $k = k_f + k_b$ \cite{Chandler78,Chandler87,Gray87}, is
actually obtained by computing the average escape time (lifetime) 
for an ensemble of trajectories in the CNH well, while somewhat arbitrarily
omitting from consideration short-lived trajectories with lifetimes $< 1500$ a.u.\ \cite{Tang94}.

The value of the mean lifetime computed by TJZR at $E=-11.5$ is close to the mean gap time 
($0.174$ ps) computed at $E=-11.88$ eV \cite{WaalkensBurbanksWiggins05}.  
Both approximate rate theories applied to the problem by TJZR give 
isomerization rates within factors of 2-3 of the trajectory values.
These approximate statistical theories 
are however only capable of describing short time kinetics \cite{Dumont86,Gray87}; 
the slow decays apparent from the trajectory data of TJZR 
(see Figure 7 of \cite{Tang94}; cf.\ Figs \ref{lifetime_all}, \ref{lifetime_1} and 
\ref{lifetime_2} of the present paper) are not 
predicted by 2- or 3-state statistical models.  The algebraic decay observed at intermediate 
times in the present work suggests that incorporation of a small number of 
approximate intramolecular bottlenecks into the statistical model \cite{Davis85,Gray86a,Gray87,Zhao92a,Rice96} 
is unlikely to lead to an accurate description of decay dynamics at longer times.

\subsection{Relation to nonstatistical behavior in reactions of organic molecules}

Finally, it is very interesting to note that the dynamical behavior associated with 
the gap time distribution analyzed here for HCN isomerization provides an 
exemplary instance of the highly nonstatistical dynamics discussed by 
Carpenter in the context of reactions of organic molecules \cite{Carpenter03,Carpenter05}.

Thus, as discussed above,  an ensemble of trajectories is launched into a reactant region 
(the HCN well)
for which there are two equivalent exits (``products'').
Any statistical theory predicts a 1:1 branching ratio, as there
is by symmetry equal probability of leaving via exit 1 or 2.
The trajectory calculations however show that a significant 
fraction of the initial ensemble of trajectories simply passes 
through the HCN well and exits directly via channel 2 in a very short time; the remainder of
the ensemble then sloshes back and forth in the well 
leading to the pulses seen at short times in the gap time distribution;
these pulses are associated with bursts of exiting trajectories alternating between
channels 2 and 1. At longer times,  $t \gtrsim 0.5$ ps,  we enter a regime in which
decay into either channel is equally likely.  By this time, however, most of the
trajectories have already exited the well.

\newpage

\section{Summary and conclusion}
\label{sec:summary}

In this paper we have revisited the general approach to classical unimolecular reaction
rates due to Thiele \cite{Thiele62} in light of recent advances in the
phase space formulation of transition state theory for multidimensional systems
\cite{Wiggins90,wwju,ujpyw,WaalkensBurbanksWiggins04,WaalkensWiggins04,WaalkensBurbanksWigginsb04,
WaalkensBurbanksWiggins05,WaalkensBurbanksWiggins05c,SchubertWaalkensWiggins06,WaalkensSchubertWiggins08}.
We showed that key concepts in Thieles's approach, namely the phase
space dividing surface separating reactants from products, the average gap time, and the
volume of phase space associated with reactive (interesting \cite{Slater59}; crossing \cite{Berne82})
trajectories, are both rigorously defined and readily computed within the
phase space approach 
\cite{WaalkensWiggins04,WaalkensBurbanksWiggins05,WaalkensBurbanksWiggins05c,WaalkensSchubertWiggins08}.

The distribution of gap times is a central element of Thiele's approach \cite{Thiele62,Dumont86}. 
Here, we have analyzed in detail the gap time distribution and associated reactant lifetime
distribution for the isomerization reaction HCN $\rightleftharpoons$ CNH, previously
studied using the methods of phase space transition state theory in 
refs \onlinecite{WaalkensBurbanksWigginsb04,WaalkensBurbanksWiggins05}.
Both algebraic (power law) and exponential decay regimes have been identified.
The dynamical origins of power law behavior or `fractional dynamics' 
in multimode Hamiltonian systems, especially at intermediate times,
remain obscure \cite{Zaslavsky05}.  
When combined with the normal form algorithm for 
computation and sampling of the phase space dividing surface, 
the cumulative gap time distribution is nevertheless a powerful diagnostic for reactive dynamics
in multi-dimensional systems.

We have also compared statistical estimates of the isomerization rate 
\cite{Thiele62,Dumont86,Berne82,WaalkensBurbanksWiggins04,WaalkensBurbanksWiggins05} with the
numerically determined decay rate.  We have found that, as noted by others \cite{Berne82}, 
correcting the RRKM estimate to account for the measure of the reactant phase space region
occupied by trapped trajectories results in a drastic overestimate of the isomerization
rate.  Compensating but as yet poorly understood trapping mechanisms in the reactant region 
serve to slow the escape rate sufficiently that the uncorrected
RRKM estimate turns out to be reasonably accurate, at least at the particular energy
studied.

In the planar model of HCN isomerization studied here, trajectories can exit the 
HCN well through either of 2 channels,
where the channels are related by a reflection symmetry.
Analysis of the decay properties of the subsensembles of trajectories that exit through  
particular channels shows that, despite a highly asymmetric distribution of initial conditions,
the complete trajectory ensemble effectively attains the full symmetry of the system phase space on
a short timescale $t \lesssim 0.5$ ps, after which the product 
branching ratio is 1:1, the  ``statistical'' value.
However, at intermediate times, this statistical product ratio is accompanied by 
\emph{nonexponential} (algebraic) decay.

We have also pointed out the close parallels between the dynamical behavior inferred from the 
gap time distribution for HCN and nonstatistical dynamics 
in reactions of organic molecules discussed by Carpenter  \cite{Carpenter03,Carpenter05}.

\acknowledgments  

H.W. acknowledges EPSRC for support under grant number  EP/E024629/1.
S.W.  acknowledges the support of the  Office of Naval Research Grant  
No.~N00014-01-1-0769.

\appendix

\section{Division of reactant phase space into trapped and reactive components}
\label{app:recurrence}

The Poincar\'e recurrence theorem \cite{Birkhoff31,Arnold78}, used in conjunction
with the geometrical structures constructed in the phase space
reaction rate theory described in Section \ref{sec:unimolecular}, allows us to
give a precise treatment of the division of the reactant phase 
space into reactive (``crossing'')  and trapped regions.

Suppose that, for a fixed energy $E$, there are $d$ saddles associated with dividing 
surfaces DS$_{i}(E)$, $i=1,\ldots,d$, where each dividing surface is divided into 
two $2n-2$ dimensional hemispheres
DS$_{i,\text{in}}(E)$ and  DS$_{i,\text{out}}(E)$ that control entrance and exit to a \emph{compact} reactant region.
We first show that every trajectory (except for a set of measure zero) that 
enters the reactant region through a dividing surface DS$_{i,\text{in}}(E)$ 
will exit the reactant region at a later time. 

To this end, consider a set 
$\overline{V}$ of $(\bar{q}, \bar{p})$ in DS$_{i,\text{in}}(E)$ of 
positive volume with respect to the (Lebesgue) measure 
$\text{d}\sigma=\text{d}^{n-1} \bar{q} \text{d}^{n-1}\bar{p}$. 
Suppose that the points in $\overline{V}$ as a set of initial conditions for 
Hamilton's equation give trajectories which stay in the reactants region for all time $\psi>0$.
Then the region swept out by these trajectories has infinite volume with respect 
to the (Lebesgue) measure $\text{d}\sigma \wedge \text{d}\psi$ (see Sec.\ \ref{subsec:volumes}). 
This contradicts the compactness of the reactant region. 
We can thus conclude that every initial condition on DS$_{i,\text{in}}(E)$ 
(except for a set of measure zero with respect to the measure 
$\text{d}\sigma=\text{d}^{n-1} \bar{q} \text{d}^{n-1}\bar{p}$) 
gives a trajectory which leaves the reactant region at a later time.
This conclusion holds for both compact and noncompact (dissociative) product regions.

In the case that the product regions are also compact, we 
can invoke the Poincar\'e recurrence theorem for Hamiltonian 
dynamics on compact energy surfaces \cite{Arnold78}:
{\em 
Consider any open set in a compact energy surface. Then, with
the possible exception of a set of (Lebesgue) measure zero, trajectories of Hamilton's equations with
initial conditions starting in this open set return infinitely often to
this set. }

Consider $\text{DS}_{i, \text{in}}(E)$, for any $i$. Trajectories
starting on this surface must enter the reactant region. That is, 
the vector field defined by Hamilton's equations,
evaluated on $\text{DS}_{i, \text{in}}(E)$, is 
transverse to $\text{DS}_{i, \text{in}}(E)$ and  pointing 
strictly into the reactant region (as proved in \cite{ujpyw}). 
This is the mathematical manifestation of the ``no-recrossing'' property.
Since the vector field defined by Hamilton's equations points
strictly into the reactant region on $\text{DS}_{i, \text{in}}(E)$ we can
construct a ``thin'' open set, ${\cal O}_{i; \text{in}}(E)$,
containing $\text{DS}_{i, \text{in}}(E)$ having the property that all
trajectories starting in this open set also enter the region.

Now, by construction, no trajectory leaving  ${\cal O}_{i; \text{in}}(E)$
and entering the reactant region can ever intersect  ${\cal O}_{i; \text{in}}(E)$
{\em without} first leaving the region (which must occur through
$\text{DS}_{j, \text{out}}(E)$, for some $j$). The reason for this is that
the vector field defined by Hamilton's equations restricted to
${\cal O}_{i; \text{in}} (E)$ is pointing strictly into the reactant region. 
By the Poincar\'e recurrence theorem, with the possible exception of a set
of zero Lebesgue measure, every trajectory starting on ${\cal O}_{i; \text{in}} (E)$ 
intersects ${\cal O}_{i; \text{in}}(E)$ infinitely often.
Therefore, we conclude that, with the possible exception of a set of zero
Lebesgue measure,  every trajectory that enters the reactant region 
exits and re-enters the region 
an infinite number of times. We can summarize as follows:
{\em Almost all trajectories that enter a given reactant region of phase space 
exit the region at a later time. Moreover, after exiting, they will
re-enter the same region at a later time, and this ``entrance-exit''
behaviour continues for all time thereafter.}

We can state this result also in a slightly different,
but equivalent, way:
{\em Almost all trajectories that exit given reactant region
will return to the same region at a later time. Moreover,
after returning, they will exit the region again at a later time,
and this ``exit-return'' behaviour continues for all time
thereafter.}

The immediate implication is that, with the  possible exception of a set of 
Lebesgue measure zero, no trajectory can escape the reactant region of phase space 
that is {\em not} in the reactive (crossing) volume.  The further implication is that, 
with the possible exception of a set of Lebesgue measure zero, 
the volume of the reactant region of phase space 
consists of two components-- the reactive volume and the trapped volume. 
The boundary between these two sets is liable to be exceedingly complicated (fractal \cite{Ott02}). 

\section{Derivation of the delay differential equation for $\bbP(t)$}
\label{app:DLGM}

In this Appendix we give a concise derivation of the delay differential equation
satisfied by the reactant lifetime distribution $\bbP(t)$ 
in the DLGM of Dumont and Brumer \cite{Dumont86}.  We consider a fixed value
of the energy $E$.

In the notation of Sec.\ \ref{subsec:gaps},
the fraction of trajectories initiated on $\text{DS}_{\text{in}}$ with gap time $s \geq t'$ is
\begin{equation}
\Int{s}{t'}{+\infty} \calP(s) = \sbar \; \bbP(t'),
\end{equation}
so that the probability of a trajectory initiated on $\text{DS}_{\text{in}}$
having a gap time $t+t' \leq s \leq t + t' + \rmd t$, $t \geq 0$, \emph{given} that $s \geq t'$ is
\begin{equation}
\frac{\calP(t+t') \rmd t}{\sbar \; \bbP(t')}
\end{equation}
The condition for statistical decay given by Dumont and Brumer can then be written \cite{Dumont86} 
\begin{equation}
\label{stat_1}
\frac{\calP(t+t')}{\sbar \; \bbP(t')} = \bbP(t), \;\; \forall t \geq 0, t' \geq \tau
\end{equation}
where $\tau$ is the \emph{relaxation time}.
The meaning of this condition is as follows:  consider those trajectories on $\text{DS}_{\text{in}}$
with gap time $s \geq t'$; the fraction of these trajectories
having gap times $t+t' \leq s \leq t + t' + \rmd t$ is equal to $\bbP(t) \rmd t$,
the fraction of reactant phase points with lifetimes $t \to t+\rmd t$, for $t' \geq \tau$.
An equivalent formulation of the condition for statistical decay is
\begin{equation}
\label{stat_2}
\frac{\rmd}{\rmd t}\, \bbP(t+t') = - \bbP(t)\bbP(t')  \;\; \forall t \geq 0, t' \geq \tau.
\end{equation}
Conditions \eqref{stat_1} and \eqref{stat_2} 
are clearly satisfied in the case of an exponential lifetime distribution,
$\bbP(t) = \Bbbk e^{-\Bbbk t}$.

To obtain the Delayed Lifetime Gap Model \cite{Dumont86} for the lifetime distribution:  

\begin{itemize}

\item Consider the gap time distribution $\calP(t)$ for the \emph{statistical} component of
reactive phase space \emph{only}.  The reactive phase space has to be partitioned into a direct and
a statistical component, perhaps using a trajectory divergence criterion \cite{Dumont86}.

\item Set $k_{\text{S}} \equiv  (\sbar)^{-1}$, where the mean gap time is evaluated by averaging
over the statistical component only.  We therefore have $\bbP(0) = k_{\text{S}}$

\item Assume that there are no gaps in the statistical component 
for $s < \tau$.  That is, $\calP(s) = 0$ for $s < \tau$ 
which implies $\bbP(t) = k_{\text{S}}$, $0 \leq t \leq \tau$.

\item  Set $t' = \tau$ (the relaxation time) to obtain the delay differential equation 
for the DLGM lifetime distribution:
\begin{equation}
\frac{\rmd}{\rmd t}\, \bbP(t+\tau) = - k_{\text{S}}\, \bbP(t) \;\; \forall t\geq 0
\end{equation}
This equation can be solved using the Laplace-Fourier transform \cite{Dumont86}.

\end{itemize}

%%  Bibliography

\def\cprime{$'$}

\newpage

\section*{Figure captions}

\begin{figure}[H]
 \caption{\label{phase_space_1} Phase space structures for unimolecular reaction (schematic).
 (a)  Definition of reactant region, NHIM  and dividing surface
 $\text{DS}(E) = \text{DS}_{\text{in}}(E) \cup \text{DS}_{\text{out}}(E)$.
 (b)  Definition of gap time $s$ and lifetime $t$.}
 \end{figure}

%\newpage

\begin{figure}[H]
 \caption{\label{hcn_pes}  Isopotential surfaces of the HCN potential
energy surface of ref.\ \onlinecite{MCH} in polar representation of the Jacobi
coordinates $r$, $R$, and $\gamma$.}
 \end{figure}

%  \newpage
   
   \begin{figure}[H]
    \caption{\label{gap_fig}  HCN gap time distribution $\calP(s)$.
    (a) Complete ensemble.  (b) Subensemble reacting via channel 1.  
     (c) Subensemble reacting via channel 2.}
   \end{figure}

% \newpage

   \begin{figure}[H]
     \caption{\label{short_life_fig}  HCN cumulative gap time (reactant lifetime)
     distribution $F(t)$ at short times.  $\text{Log} F(t)$ is plotted vs $t$ for
     $0 \leq t \leq 0.4$ ps. (a)  Total ensemble.
     (b) Subensemble reacting via channel 1.  
     (c) Subensemble reacting via channel 2.}
    \end{figure}

% \newpage
 
 \begin{figure}[H]
  \caption{\label{lifetime_all} Cumulative gap time (reactant lifetime)
  distribution $F(t)$ for the complete ensemble.
  (a)  A log-log plot shows power law decay at intermediate times. 
  (b)  Log plot shows exponential decay ($10 \leq t \leq 20$ ps).}
  \end{figure} 
 
%\newpage
 
 \begin{figure}[H]
  \caption{\label{lifetime_1} Cumulative gap time (reactant lifetime)
  distribution $F(t)$ for the subensemble reacting via channel 1.
  (a)  A log-log plot shows power law decay at intermediate times. 
  (b)  Log plot shows exponential decay ($10 \leq t \leq 20$ ps).}
  \end{figure}

%\newpage
 
 \begin{figure}[H]
  \caption{\label{lifetime_2} Cumulative gap time (reactant lifetime)
  distribution $F(t)$ for the subensemble reacting via channel 2.
  (a)  A log-log plot shows power law decay at intermediate times. 
  (b)  Log plot shows exponential decay ($10 \leq t \leq 20$ ps).}
  \end{figure}

% \newpage
  
  \begin{figure}[H]
   \caption{\label{survival}  HCN survival probability $P_{\text{S}}(t)$.   
   $P_{\text{S}}(t)$ is the fraction of an ensemble of
   trajectories uniformly distributed throughout the HCN region of phase space at $t=0$ remaining in the
   well at time $t$.  Trajectories are removed from the ensemble once they exit the HCN region by 
   crossing $\text{DS}_{j, \text{out}}$, $j=1,2$; 
   they cannot re-enter the region.}
  \end{figure}

% \newpage
  
  \begin{figure}[H]
   \caption{\label{nlog12}  The log of the number $N(t)$ of trajectories remaining at time
   $t$ versus $t$ is plotted for each subensemble: channel 1 (blue), channel 2 (red).
   }
  \end{figure}

%%  Figures

\newpage

\begin{figure}[H]
 \centering
 \includegraphics[width=4.5in]{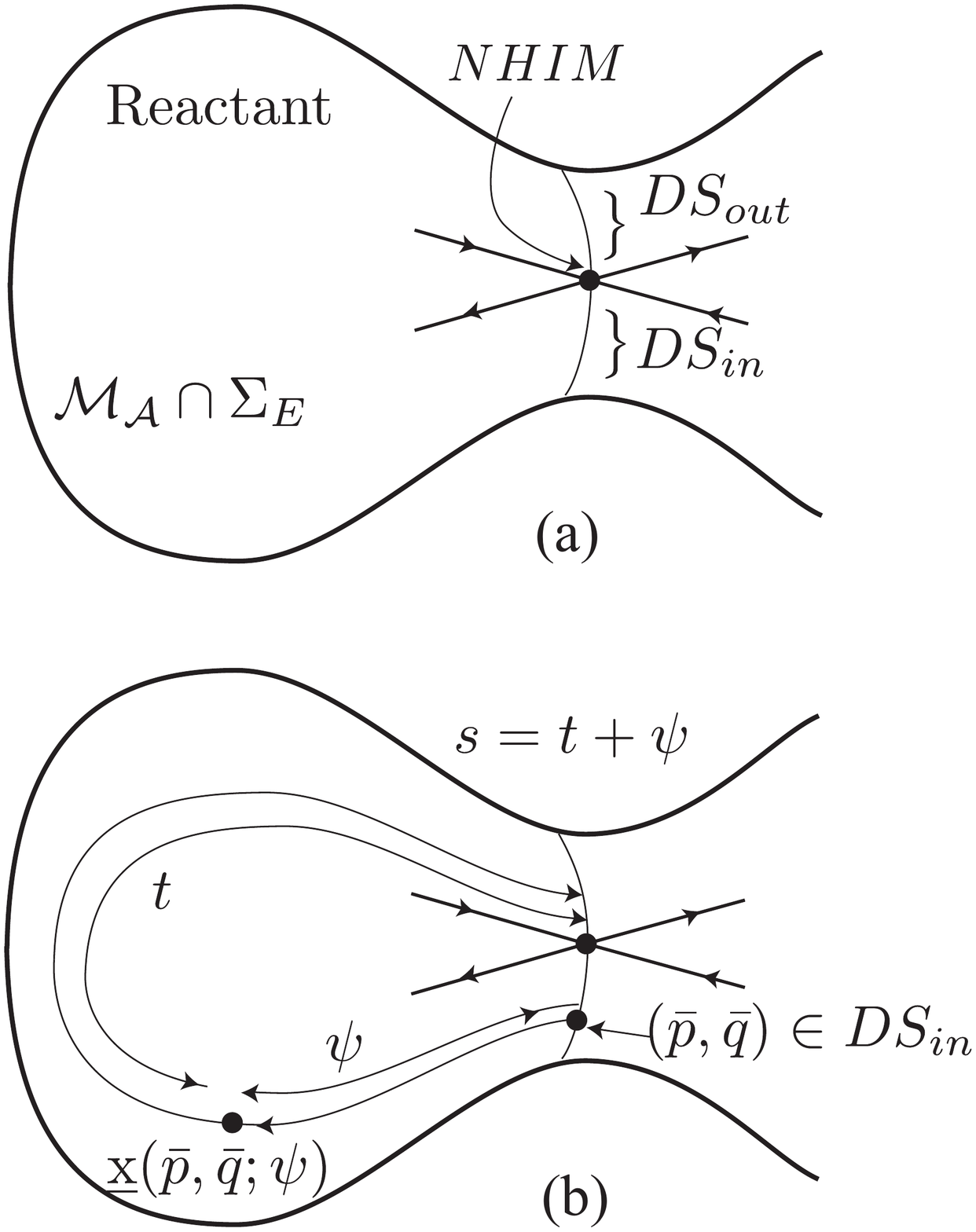}
 \vspace*{0.5in}
 \end{figure} 
 
   \vspace*{1.5cm}
   FIGURE 1

\newpage

\begin{figure}[H]
 \centering
 \includegraphics[width=5.5in]{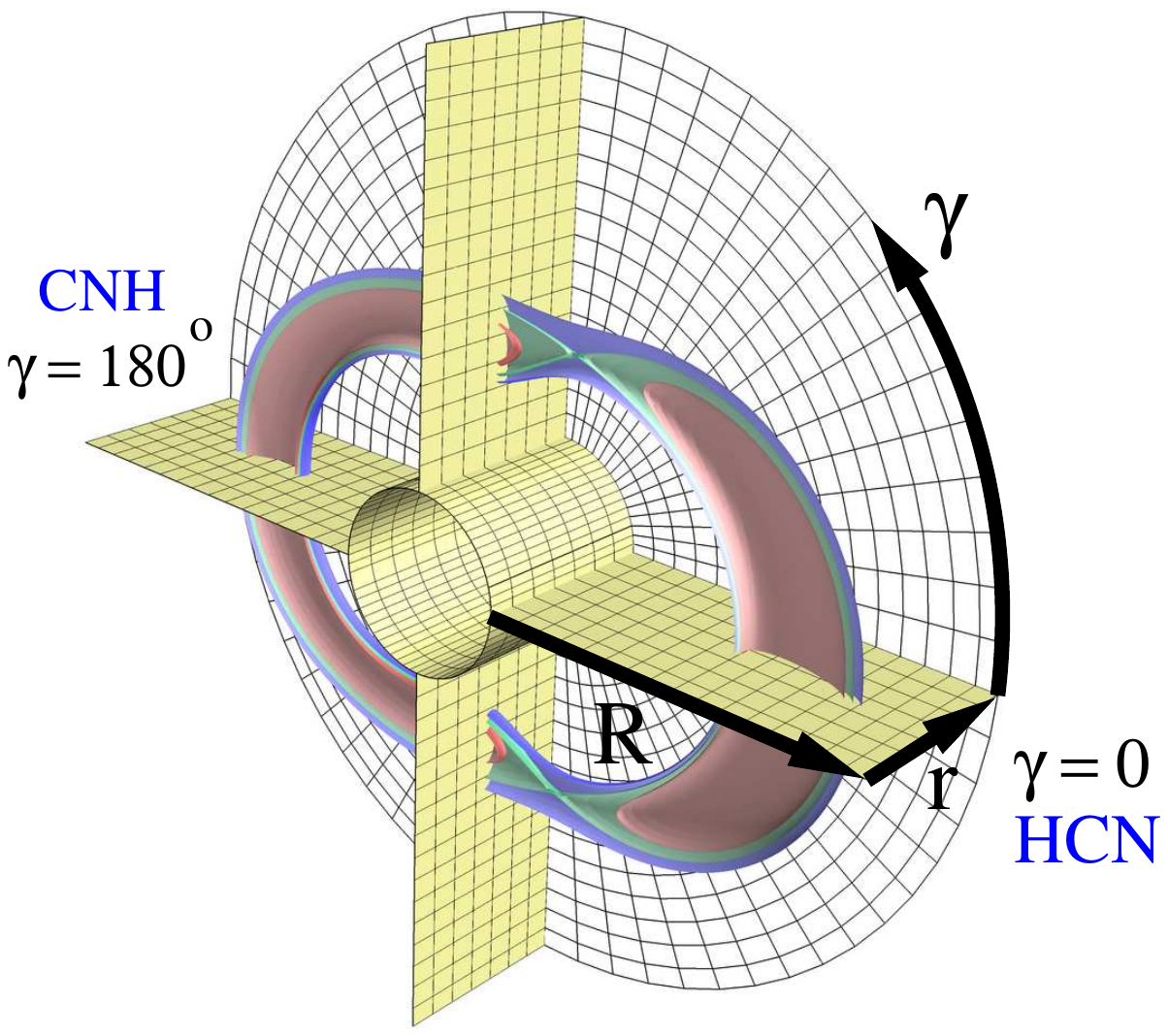}
 \end{figure} 
 
%   \vspace*{1.5cm}
   FIGURE 2

  \newpage
   
   \begin{figure}[H]
    \centering
      \includegraphics[width=6.in]{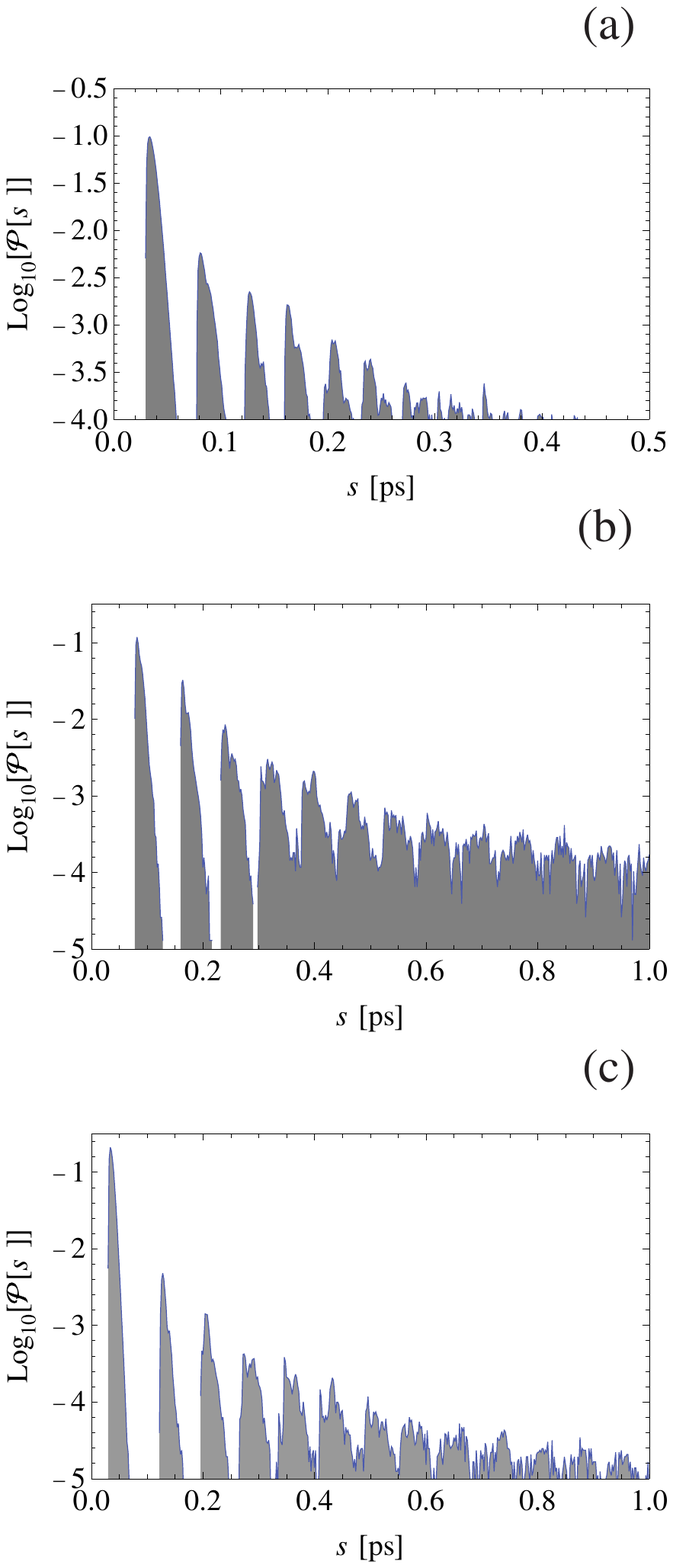}
   \end{figure} 

 \vspace*{.5cm}
   FIGURE 3

 \newpage

   \begin{figure}[H]
     \centering
     \includegraphics[width=6.in]{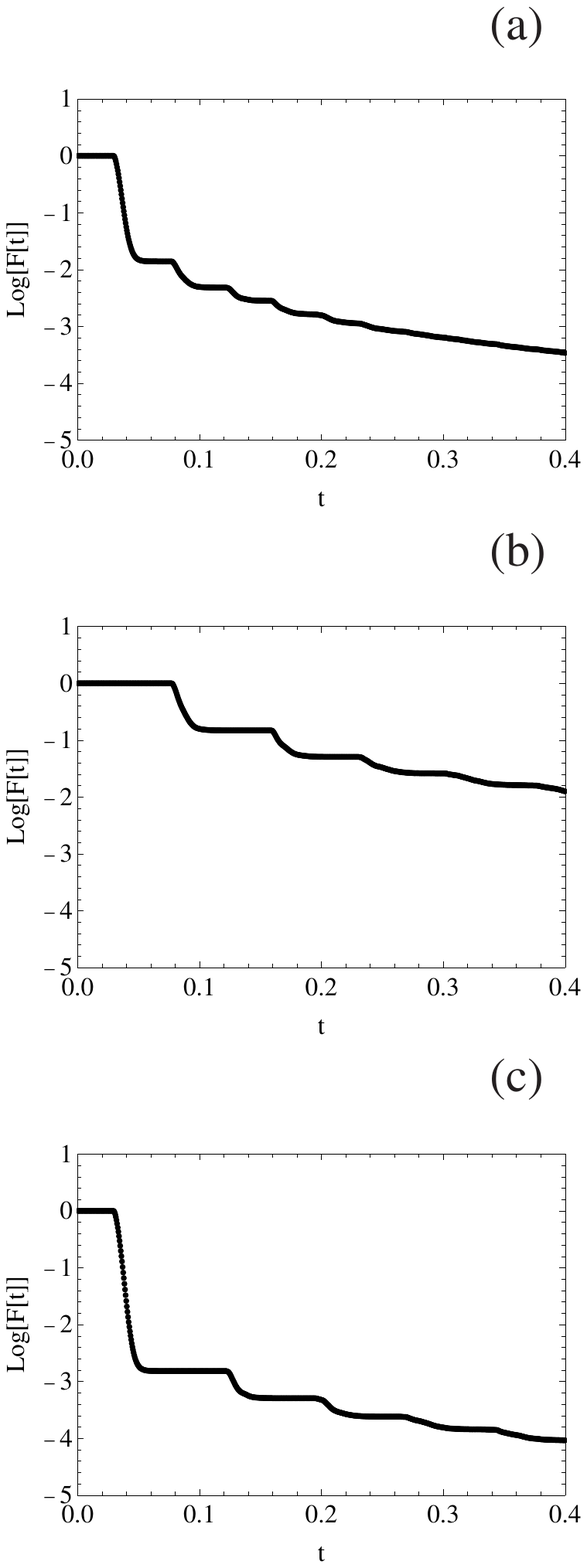}
    \end{figure} 
       
 \vspace*{0.5cm}
   FIGURE 4

 \newpage
 
 \begin{figure}[H]
  \centering
  \includegraphics[width=6.5in]{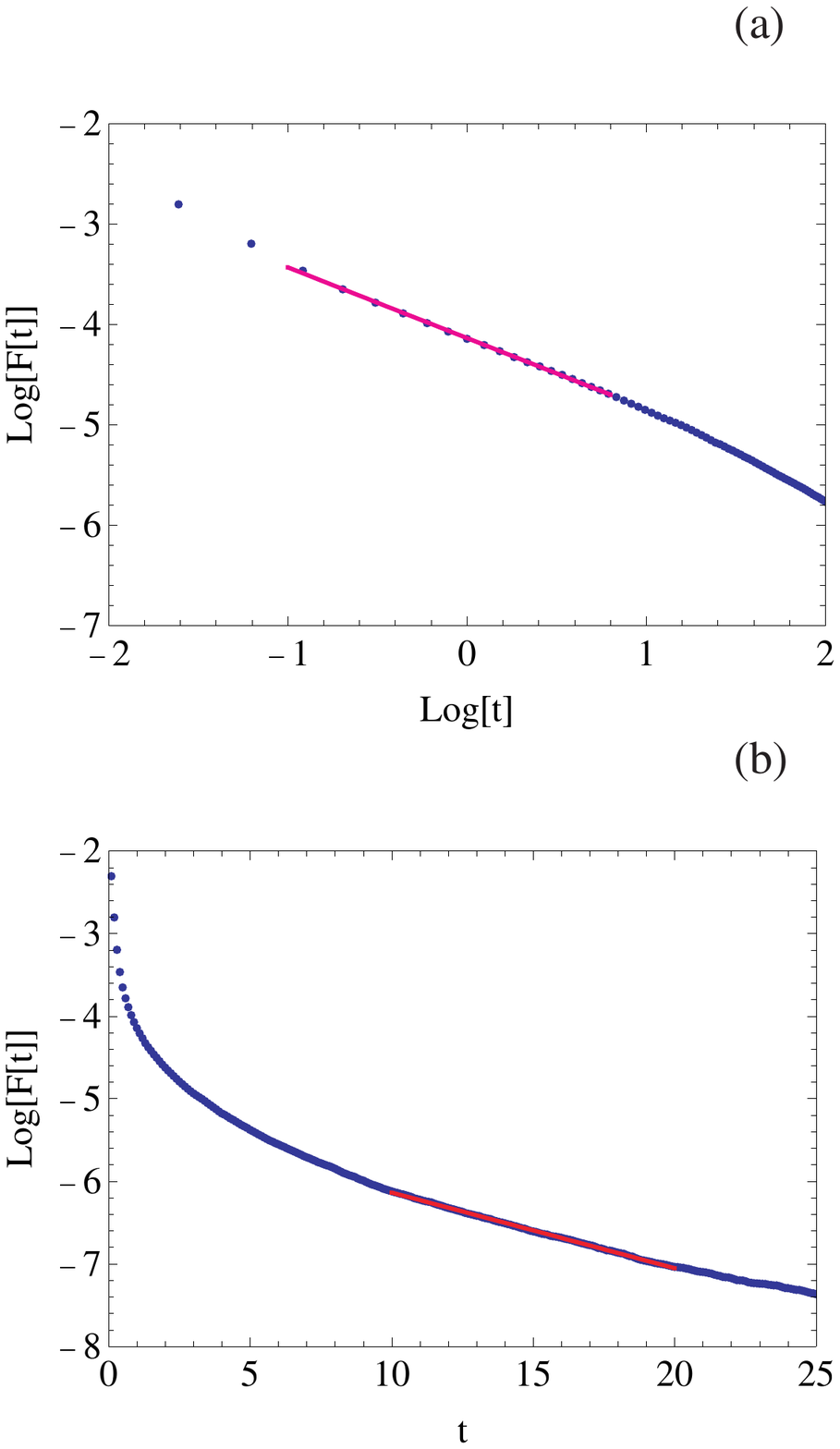}
  \end{figure} 
  
% \vspace*{1.5cm}
   FIGURE 5

\newpage
 
 \begin{figure}[H]
  \centering
  \includegraphics[width=6.5in]{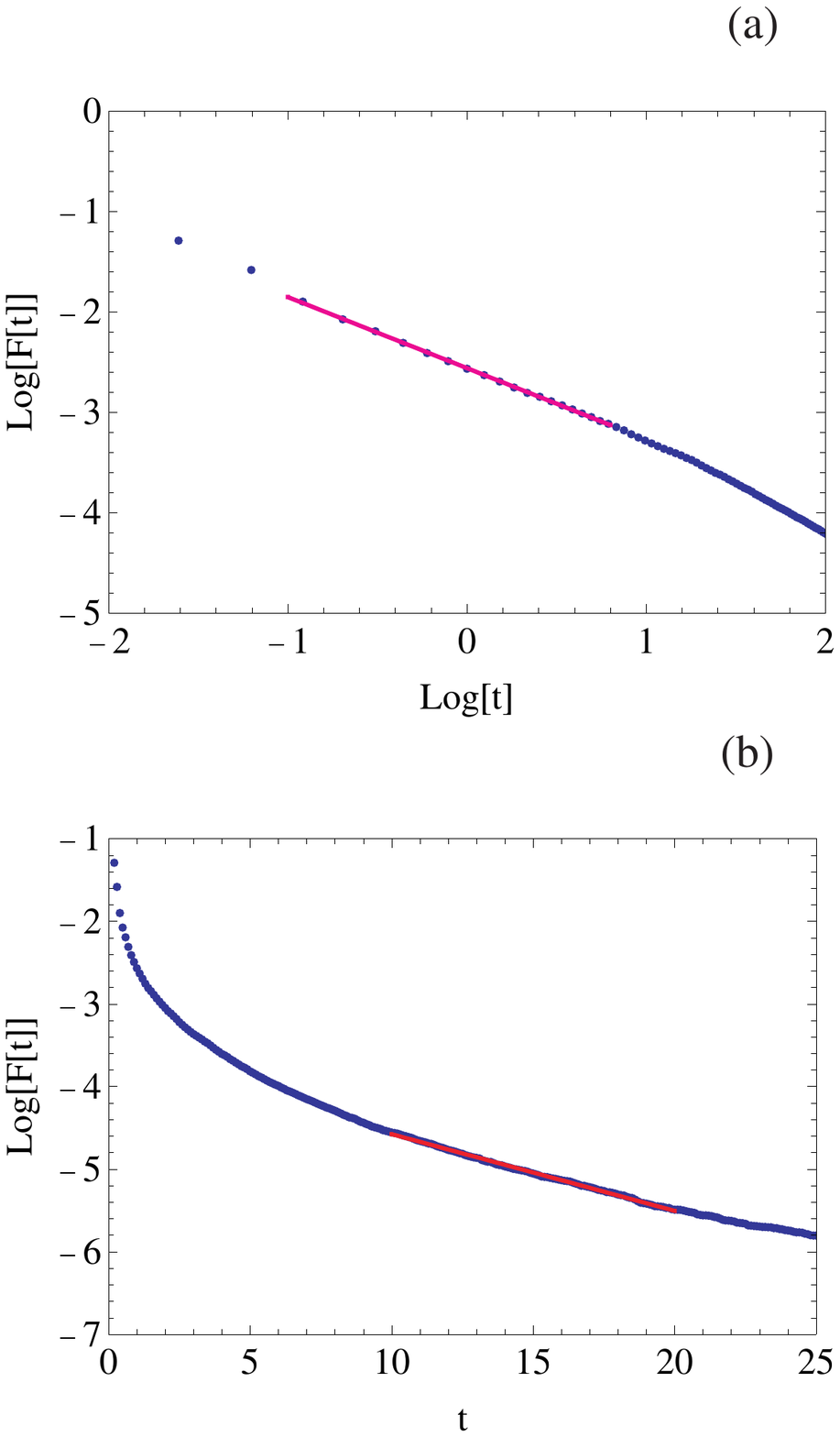}
  \end{figure}

% \vspace*{1.5cm}
   FIGURE 6

\newpage
 
 \begin{figure}[H]
  \centering
 \includegraphics[width=6.5in]{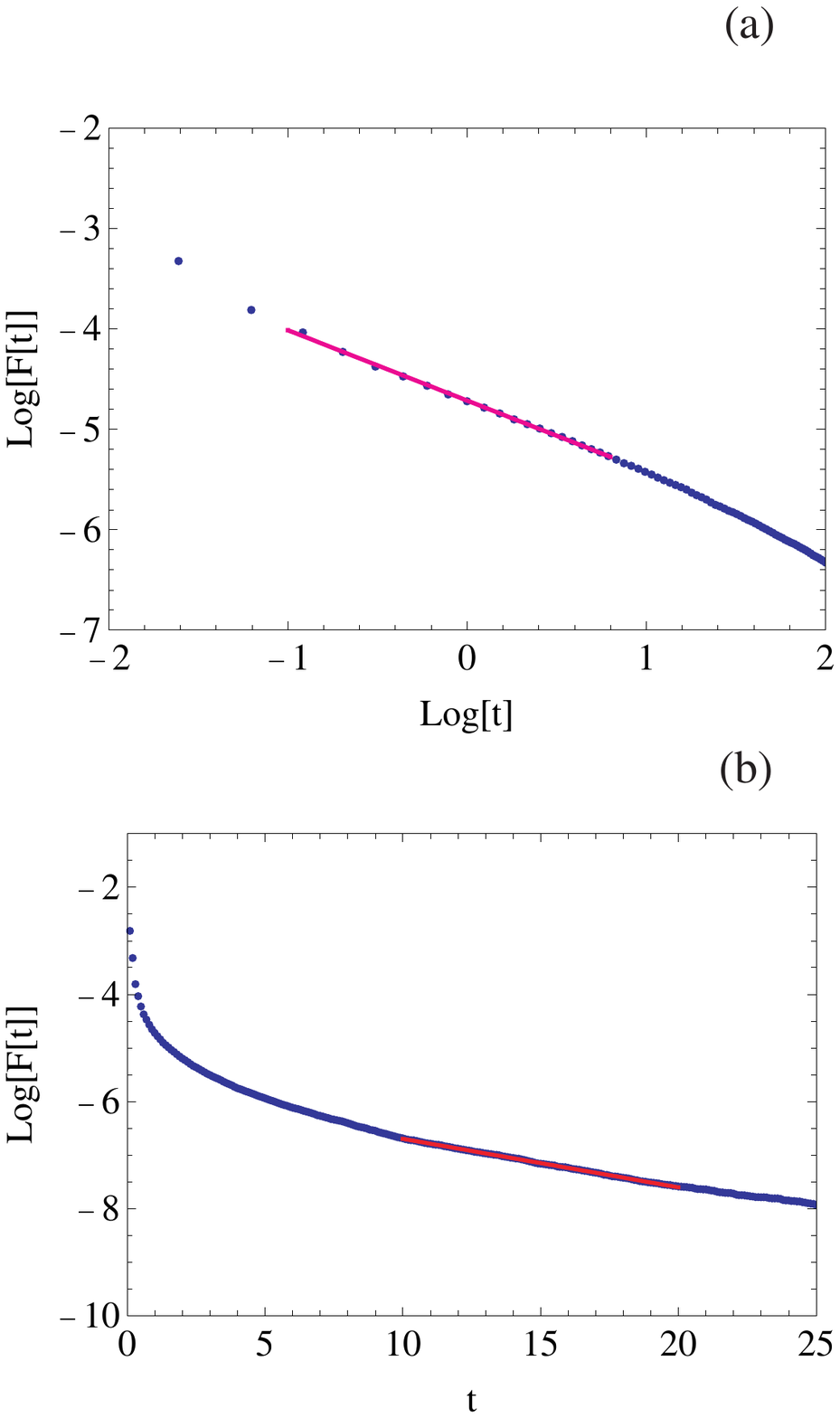}
  \end{figure}
  
% \vspace*{1.5cm}
   FIGURE 7

 \newpage
  
  \begin{figure}[H]
   \centering
   \includegraphics[width=6.5in]{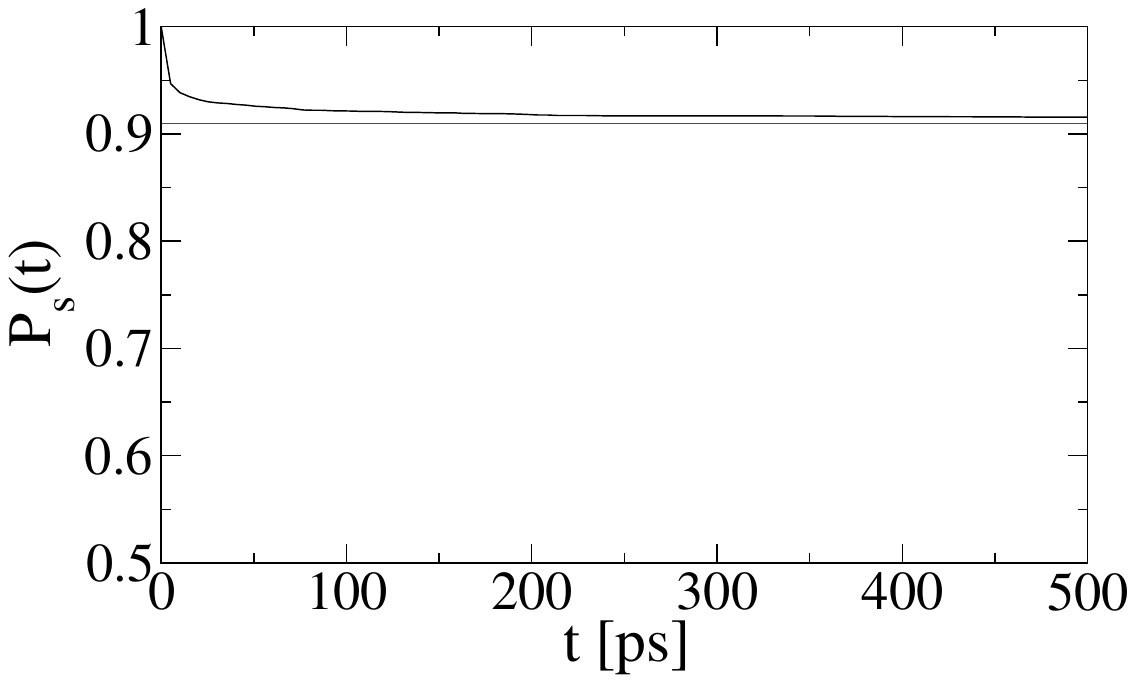}
  \end{figure} 

% \vspace*{1.5cm}
   FIGURE 8

 \newpage
  
  \begin{figure}[H]
   \centering
   \includegraphics[width=6.5in]{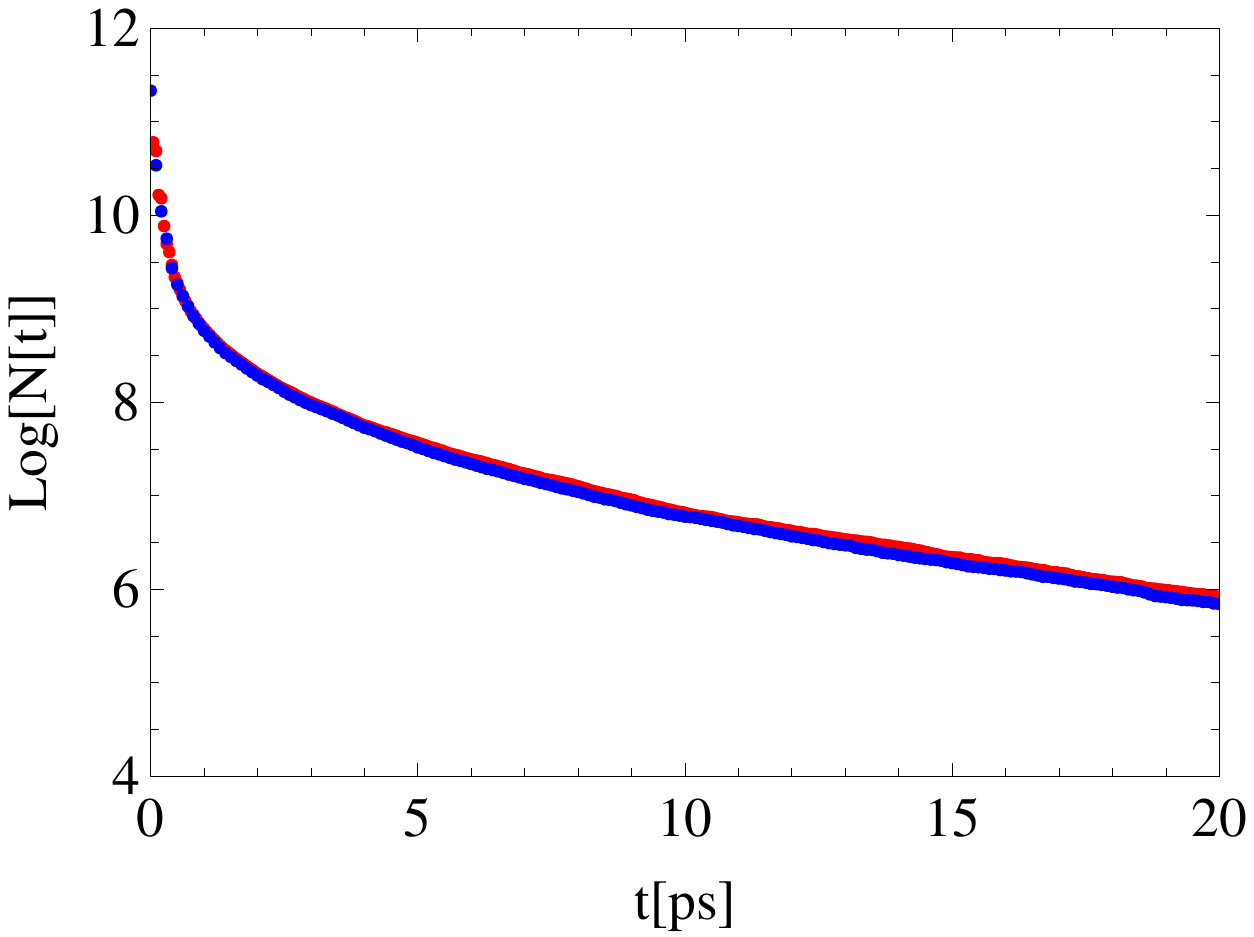}
  \end{figure}

%  \vspace*{1.5cm}
    FIGURE 9


\begin{thebibliography}{100}
\expandafter\ifx\csname bibnamefont\endcsname\relax
  \def\bibnamefont#1{#1}\fi
\expandafter\ifx\csname bibfnamefont\endcsname\relax
  \def\bibfnamefont#1{#1}\fi
\expandafter\ifx\csname url\endcsname\relax
  \def\url#1{\texttt{#1}}\fi
\expandafter\ifx\csname urlprefix\endcsname\relax\def\urlprefix{URL }\fi
\expandafter\ifx\csname bibinfo\endcsname\relax \def\bibinfo#1#2{#2}\fi
\expandafter\ifx\csname eprint\endcsname\relax \def\eprint#1{#1}\fi

\bibitem{Marcelin15}
\bibinfo{author}{\bibfnamefont{R.}~\bibnamefont{Marcelin}},
  \bibinfo{journal}{Annales de Physique} \textbf{\bibinfo{volume}{3}},
  \bibinfo{pages}{120} (\bibinfo{year}{1915}).

\bibitem{Rice27}
\bibinfo{author}{\bibfnamefont{O.~K.} \bibnamefont{Rice}} \bibnamefont{and}
  \bibinfo{author}{\bibfnamefont{H.~C.} \bibnamefont{Ramsperger}},
  \bibinfo{journal}{J. Am. Chem. Soc.} \textbf{\bibinfo{volume}{49}},
  \bibinfo{pages}{1617} (\bibinfo{year}{1927}).

\bibitem{Kassel28}
\bibinfo{author}{\bibfnamefont{L.~S.} \bibnamefont{Kassel}},
  \bibinfo{journal}{J. Phys. Chem.} \textbf{\bibinfo{volume}{32}},
  \bibinfo{pages}{225} (\bibinfo{year}{1928}).

\bibitem{Kassel28a}
\bibinfo{author}{\bibfnamefont{L.~S.} \bibnamefont{Kassel}},
  \bibinfo{journal}{J. Phys. Chem.} \textbf{\bibinfo{volume}{32}},
  \bibinfo{pages}{1065} (\bibinfo{year}{1928}).

\bibitem{Marcus51}
\bibinfo{author}{\bibfnamefont{R.~A.} \bibnamefont{Marcus}} \bibnamefont{and}
  \bibinfo{author}{\bibfnamefont{O.~K.} \bibnamefont{Rice}},
  \bibinfo{journal}{J. Phys. Colloid Chem.} \textbf{\bibinfo{volume}{55}},
  \bibinfo{pages}{894} (\bibinfo{year}{1951}).

\bibitem{Marcus52}
\bibinfo{author}{\bibfnamefont{R.~A.} \bibnamefont{Marcus}},
  \bibinfo{journal}{J. Chem. Phys.} \textbf{\bibinfo{volume}{20}},
  \bibinfo{pages}{359} (\bibinfo{year}{1952}).

\bibitem{Kassel32}
\bibinfo{author}{\bibfnamefont{L.~S.} \bibnamefont{Kassel}},
  \emph{\bibinfo{title}{{The Kinetics of Homogenous Gas Reactions}}}
  (\bibinfo{publisher}{Chemical Catalog Company}, \bibinfo{address}{New York},
  \bibinfo{year}{1932}).

\bibitem{Hinshelwood40}
\bibinfo{author}{\bibfnamefont{C.~N.} \bibnamefont{Hinshelwood}},
  \emph{\bibinfo{title}{{The Kinetics of Chemical Change}}}
  (\bibinfo{publisher}{Clarendon Press}, \bibinfo{address}{Oxford},
  \bibinfo{year}{1940}).

\bibitem{Slater59}
\bibinfo{author}{\bibfnamefont{N.~B.} \bibnamefont{Slater}},
  \emph{\bibinfo{title}{{Theory of Unimolecular Reactions}}}
  (\bibinfo{publisher}{Cornell University Press}, \bibinfo{address}{Ithaca,
  NY}, \bibinfo{year}{1959}).

\bibitem{Bunker66}
\bibinfo{author}{\bibfnamefont{D.~L.} \bibnamefont{Bunker}},
  \emph{\bibinfo{title}{{Theory of Elementary Gas Reaction Rates}}}
  (\bibinfo{publisher}{Pergamon}, \bibinfo{address}{Oxford},
  \bibinfo{year}{1966}).

\bibitem{Johnston66}
\bibinfo{author}{\bibfnamefont{H.~S.} \bibnamefont{Johnston}},
  \emph{\bibinfo{title}{{Gas Phase Reaction Rate Theory}}}
  (\bibinfo{publisher}{Ronald Press}, \bibinfo{address}{New York},
  \bibinfo{year}{1966}).

\bibitem{Rice67}
\bibinfo{author}{\bibfnamefont{O.~K.} \bibnamefont{Rice}},
  \emph{\bibinfo{title}{{Statistical Mechanics, Thermodynamics, and Kinetics}}}
  (\bibinfo{publisher}{W. H. Freeman}, \bibinfo{address}{San Francisco},
  \bibinfo{year}{1967}).

\bibitem{Laidler69}
\bibinfo{author}{\bibfnamefont{K.~J.} \bibnamefont{Laidler}},
  \emph{\bibinfo{title}{{Theories of Chemical Reaction Rates}}}
  (\bibinfo{publisher}{McGraw-Hill}, \bibinfo{address}{New York},
  \bibinfo{year}{1969}).

\bibitem{Robinson72}
\bibinfo{author}{\bibfnamefont{P.~J.} \bibnamefont{Robinson}} \bibnamefont{and}
  \bibinfo{author}{\bibfnamefont{K.~A.} \bibnamefont{Holbrook}},
  \emph{\bibinfo{title}{{Unimolecular Reactions}}} (\bibinfo{publisher}{Wiley},
  \bibinfo{address}{New York}, \bibinfo{year}{1972}).

\bibitem{Forst73}
\bibinfo{author}{\bibfnamefont{W.}~\bibnamefont{Forst}},
  \emph{\bibinfo{title}{{Theory of Unimolecular Reactions}}}
  (\bibinfo{publisher}{Academic}, \bibinfo{address}{New York},
  \bibinfo{year}{1973}).

\bibitem{Nikitin74}
\bibinfo{author}{\bibfnamefont{E.~E.} \bibnamefont{Nikitin}},
  \emph{\bibinfo{title}{{Theory of Elementary Atomic and Molecular Processes}}}
  (\bibinfo{publisher}{Clarendon Press}, \bibinfo{address}{Oxford},
  \bibinfo{year}{1974}).

\bibitem{Smith80}
\bibinfo{author}{\bibfnamefont{I.~W.~M.} \bibnamefont{Smith}},
  \emph{\bibinfo{title}{{Kinetics and Dynamics of Elementary Gas Reactions}}}
  (\bibinfo{publisher}{Butterworths}, \bibinfo{address}{London},
  \bibinfo{year}{1980}).

\bibitem{Beynon84}
\bibinfo{author}{\bibfnamefont{J.~H.} \bibnamefont{Beynon}} \bibnamefont{and}
  \bibinfo{author}{\bibfnamefont{J.~R.} \bibnamefont{Gilbert}},
  \emph{\bibinfo{title}{{Application of Transition State Theory to Unimolecular
  Reactions: an Introduction}}} (\bibinfo{publisher}{Wiley},
  \bibinfo{address}{New York}, \bibinfo{year}{1984}).

\bibitem{Pritchard84}
\bibinfo{author}{\bibfnamefont{H.~O.} \bibnamefont{Pritchard}},
  \emph{\bibinfo{title}{{The Quantum Theory of Unimolecular Reactions}}}
  (\bibinfo{publisher}{Cambridge University Press},
  \bibinfo{address}{Cambridge}, \bibinfo{year}{1984}).

\bibitem{Wardlaw88}
\bibinfo{author}{\bibfnamefont{D.}~\bibnamefont{Wardlaw}} \bibnamefont{and}
  \bibinfo{author}{\bibfnamefont{R.~A.} \bibnamefont{Marcus}},
  \bibinfo{journal}{Adv. Chem. Phys.} \textbf{\bibinfo{volume}{70}},
  \bibinfo{pages}{231} (\bibinfo{year}{1988}).

\bibitem{Gilbert90}
\bibinfo{author}{\bibfnamefont{R.~G.} \bibnamefont{Gilbert}} \bibnamefont{and}
  \bibinfo{author}{\bibfnamefont{S.~C.} \bibnamefont{Smith}},
  \emph{\bibinfo{title}{{Theory of Unimolecular and Recombination Reactions}}}
  (\bibinfo{publisher}{Blackwell Scientific}, \bibinfo{address}{Oxford},
  \bibinfo{year}{1990}).

\bibitem{Baer96}
\bibinfo{author}{\bibfnamefont{T.}~\bibnamefont{Baer}} \bibnamefont{and}
  \bibinfo{author}{\bibfnamefont{W.~L.} \bibnamefont{Hase}},
  \emph{\bibinfo{title}{{Unimolecular Reaction Dynamics}}}
  (\bibinfo{publisher}{Oxford University Press}, \bibinfo{address}{New York},
  \bibinfo{year}{1996}).

\bibitem{Forst03}
\bibinfo{author}{\bibfnamefont{W.}~\bibnamefont{Forst}},
  \emph{\bibinfo{title}{{Unimolecular Reactions}}}
  (\bibinfo{publisher}{Cambridge University Press},
  \bibinfo{address}{Cambridge}, \bibinfo{year}{2003}).

\bibitem{Henriksen08}
\bibinfo{author}{\bibfnamefont{N.~E.} \bibnamefont{Henriksen}}
  \bibnamefont{and} \bibinfo{author}{\bibfnamefont{F.~Y.}
  \bibnamefont{Hansen}}, \emph{\bibinfo{title}{{Theories of Molecular Reaction
  Dynamics: The Microscopic Foundation of Chemical Kinetics}}}
  (\bibinfo{publisher}{Oxford University Press}, \bibinfo{address}{New York},
  \bibinfo{year}{2008}).

\bibitem{Pollak05}
\bibinfo{author}{\bibfnamefont{E.}~\bibnamefont{Pollak}} \bibnamefont{and}
  \bibinfo{author}{\bibfnamefont{P.}~\bibnamefont{Talkner}},
  \bibinfo{journal}{CHAOS} \textbf{\bibinfo{volume}{15}},
  \bibinfo{pages}{026116} (\bibinfo{year}{2005}).

\bibitem{Slater56}
\bibinfo{author}{\bibfnamefont{N.~B.} \bibnamefont{Slater}},
  \bibinfo{journal}{J. Chem. Phys.}
  \textbf{\bibinfo{volume}{24}}(\bibinfo{number}{6}), \bibinfo{pages}{1256}
  (\bibinfo{year}{1956}).

\bibitem{Thiele62}
\bibinfo{author}{\bibfnamefont{E.}~\bibnamefont{Thiele}}, \bibinfo{journal}{J.
  Chem. Phys.} \textbf{\bibinfo{volume}{36}}(\bibinfo{number}{6}),
  \bibinfo{pages}{1466} (\bibinfo{year}{1962}).

\bibitem{Bunker73}
\bibinfo{author}{\bibfnamefont{D.}~\bibnamefont{Bunker}} \bibnamefont{and}
  \bibinfo{author}{\bibfnamefont{W.~L.} \bibnamefont{Hase}},
  \bibinfo{journal}{J. Chem. Phys.} \textbf{\bibinfo{volume}{59}},
  \bibinfo{pages}{4621} (\bibinfo{year}{1973}).

\bibitem{Dumont86}
\bibinfo{author}{\bibfnamefont{R.~S.} \bibnamefont{Dumont}} \bibnamefont{and}
  \bibinfo{author}{\bibfnamefont{P.}~\bibnamefont{Brumer}},
  \bibinfo{journal}{J. Phys. Chem.} \textbf{\bibinfo{volume}{90}},
  \bibinfo{pages}{3509} (\bibinfo{year}{1986}).

\bibitem{Dumont92}
\bibinfo{author}{\bibfnamefont{R.~S.} \bibnamefont{Dumont}} \bibnamefont{and}
  \bibinfo{author}{\bibfnamefont{P.}~\bibnamefont{Brumer}},
  \bibinfo{journal}{Chem. Phys. Lett.} \textbf{\bibinfo{volume}{188}},
  \bibinfo{pages}{565} (\bibinfo{year}{1992}).

\bibitem{Dumont89}
\bibinfo{author}{\bibfnamefont{R.~S.} \bibnamefont{Dumont}},
  \bibinfo{journal}{J. Chem. Phys.} \textbf{\bibinfo{volume}{91}},
  \bibinfo{pages}{4679} (\bibinfo{year}{1989}).

\bibitem{Dumont89a}
\bibinfo{author}{\bibfnamefont{R.~S.} \bibnamefont{Dumont}},
  \bibinfo{journal}{J. Chem. Phys.} \textbf{\bibinfo{volume}{91}},
  \bibinfo{pages}{6839} (\bibinfo{year}{1989}).

\bibitem{Dumont92a}
\bibinfo{author}{\bibfnamefont{R.~S.} \bibnamefont{Dumont}} \bibnamefont{and}
  \bibinfo{author}{\bibfnamefont{S.}~\bibnamefont{Jain}}, \bibinfo{journal}{J.
  Chem. Phys.} \textbf{\bibinfo{volume}{97}}, \bibinfo{pages}{1227}
  (\bibinfo{year}{1992}).

\bibitem{Jain93}
\bibinfo{author}{\bibfnamefont{S.}~\bibnamefont{Jain}},
  \bibinfo{author}{\bibfnamefont{S.}~\bibnamefont{Bleher}}, \bibnamefont{and}
  \bibinfo{author}{\bibfnamefont{R.~S.} \bibnamefont{Dumont}},
  \bibinfo{journal}{J. Chem. Phys.} \textbf{\bibinfo{volume}{99}},
  \bibinfo{pages}{7793} (\bibinfo{year}{1993}).

\bibitem{Brumer80}
\bibinfo{author}{\bibfnamefont{P.}~\bibnamefont{Brumer}},
  \bibinfo{author}{\bibfnamefont{D.~E.} \bibnamefont{Fitz}}, \bibnamefont{and}
  \bibinfo{author}{\bibfnamefont{D.}~\bibnamefont{Wardlaw}},
  \bibinfo{journal}{J. Chem. Phys.}
  \textbf{\bibinfo{volume}{72}}(\bibinfo{number}{1}), \bibinfo{pages}{386}
  (\bibinfo{year}{1980}).

\bibitem{Pollak81}
\bibinfo{author}{\bibfnamefont{E.}~\bibnamefont{Pollak}}, \bibinfo{journal}{J.
  Chem. Phys.} \textbf{\bibinfo{volume}{74}}, \bibinfo{pages}{6763}
  (\bibinfo{year}{1981}).

\bibitem{Tabor81}
\bibinfo{author}{\bibfnamefont{M.}~\bibnamefont{Tabor}}, \bibinfo{journal}{Adv.
  Chem. Phys.} \textbf{\bibinfo{volume}{XLVI}}, \bibinfo{pages}{73}
  (\bibinfo{year}{1981}).

\bibitem{Rice81}
\bibinfo{author}{\bibfnamefont{S.~A.} \bibnamefont{Rice}},
  \bibinfo{journal}{Adv. Chem. Phys.} \textbf{\bibinfo{volume}{XLVII}},
  \bibinfo{pages}{117} (\bibinfo{year}{1981}).

\bibitem{Brumer81}
\bibinfo{author}{\bibfnamefont{P.}~\bibnamefont{Brumer}},
  \bibinfo{journal}{Adv. Chem. Phys.} \textbf{\bibinfo{volume}{XLVII}},
  \bibinfo{pages}{201} (\bibinfo{year}{1981}).

\bibitem{Brumer88}
\bibinfo{author}{\bibfnamefont{P.}~\bibnamefont{Brumer}} \bibnamefont{and}
  \bibinfo{author}{\bibfnamefont{M.}~\bibnamefont{Shapiro}},
  \bibinfo{journal}{Adv. Chem. Phys.} \textbf{\bibinfo{volume}{70}},
  \bibinfo{pages}{365} (\bibinfo{year}{1988}).

\bibitem{Nordholm75}
\bibinfo{author}{\bibfnamefont{S.}~\bibnamefont{Nordholm}} \bibnamefont{and}
  \bibinfo{author}{\bibfnamefont{S.~A.} \bibnamefont{Rice}},
  \bibinfo{journal}{J. Chem. Phys.} \textbf{\bibinfo{volume}{62}},
  \bibinfo{pages}{157} (\bibinfo{year}{1975}).

\bibitem{Rice00}
\bibinfo{author}{\bibfnamefont{S.~A.} \bibnamefont{Rice}}, \bibinfo{journal}{J.
  Stat. Phys.} \textbf{\bibinfo{volume}{101}}, \bibinfo{pages}{187}
  (\bibinfo{year}{2000}).

\bibitem{Gong05}
\bibinfo{author}{\bibfnamefont{J.~B.} \bibnamefont{Gong}} \bibnamefont{and}
  \bibinfo{author}{\bibfnamefont{P.}~\bibnamefont{Brumer}},
  \bibinfo{journal}{Ann. Rev. Phys. Chem.} \textbf{\bibinfo{volume}{56}},
  \bibinfo{pages}{1} (\bibinfo{year}{2005}).

\bibitem{Thiele80}
\bibinfo{author}{\bibfnamefont{E.}~\bibnamefont{Thiele}},
  \bibinfo{author}{\bibfnamefont{M.~F.} \bibnamefont{Goodman}},
  \bibnamefont{and} \bibinfo{author}{\bibfnamefont{J.}~\bibnamefont{Stone}},
  \bibinfo{journal}{Opt. Eng.}
  \textbf{\bibinfo{volume}{19}}(\bibinfo{number}{6}), \bibinfo{pages}{10}
  (\bibinfo{year}{1980}).

\bibitem{Carpenter92}
\bibinfo{author}{\bibfnamefont{B.~K.} \bibnamefont{Carpenter}},
  \bibinfo{journal}{Acc. Chem. Res.} \textbf{\bibinfo{volume}{25}},
  \bibinfo{pages}{520} (\bibinfo{year}{1992}).

\bibitem{Carpenter03}
\bibinfo{author}{\bibfnamefont{B.~K.} \bibnamefont{Carpenter}},
  \bibinfo{journal}{J. Phys. Org. Chem.} \textbf{\bibinfo{volume}{16}},
  \bibinfo{pages}{858} (\bibinfo{year}{2003}).

\bibitem{Carpenter05}
\bibinfo{author}{\bibfnamefont{B.~K.} \bibnamefont{Carpenter}},
  \bibinfo{journal}{Ann. Rev. Phys. Chem.} \textbf{\bibinfo{volume}{56}},
  \bibinfo{pages}{57} (\bibinfo{year}{2005}).

\bibitem{Bach06}
\bibinfo{author}{\bibfnamefont{A.}~\bibnamefont{Bach}},
  \bibinfo{author}{\bibfnamefont{J.~M.} \bibnamefont{Hostettler}},
  \bibnamefont{and} \bibinfo{author}{\bibfnamefont{P.}~\bibnamefont{Chen}},
  \bibinfo{journal}{J. Chem. Phys.} \textbf{\bibinfo{volume}{125}},
  \bibinfo{pages}{Art. No. 024304} (\bibinfo{year}{2006}).

\bibitem{Berblinger94}
\bibinfo{author}{\bibfnamefont{M.}~\bibnamefont{Berblinger}} \bibnamefont{and}
  \bibinfo{author}{\bibfnamefont{C.}~\bibnamefont{Schlier}},
  \bibinfo{journal}{J. Chem. Phys.} \textbf{\bibinfo{volume}{101}},
  \bibinfo{pages}{4750} (\bibinfo{year}{1994}).

\bibitem{MacKay87}
\bibinfo{author}{\bibfnamefont{R.~S.} \bibnamefont{MacKay}} \bibnamefont{and}
  \bibinfo{author}{\bibfnamefont{J.~D.} \bibnamefont{Meiss}},
  \emph{\bibinfo{title}{{Hamiltonian Dynamical Systems: A reprint selection}}}
  (\bibinfo{publisher}{Taylor and Francis}, \bibinfo{address}{London},
  \bibinfo{year}{1987}).

\bibitem{Lichtenberg92}
\bibinfo{author}{\bibfnamefont{A.~J.} \bibnamefont{Lichtenberg}}
  \bibnamefont{and} \bibinfo{author}{\bibfnamefont{M.~A.}
  \bibnamefont{Lieberman}}, \emph{\bibinfo{title}{{Regular and Chaotic
  Dynamics}}} (\bibinfo{publisher}{Springer Verlag}, \bibinfo{address}{New
  York}, \bibinfo{year}{1992}), 2nd ed.

\bibitem{Wiggins92}
\bibinfo{author}{\bibfnamefont{S.}~\bibnamefont{Wiggins}},
  \emph{\bibinfo{title}{Chaotic transport in dynamical systems}}
  (\bibinfo{publisher}{Springer-Verlag}, \bibinfo{year}{1992}).

\bibitem{Arnold06}
\bibinfo{author}{\bibfnamefont{V.~I.} \bibnamefont{Arnold}},
  \bibinfo{author}{\bibfnamefont{V.~V.} \bibnamefont{Kozlov}},
  \bibnamefont{and} \bibinfo{author}{\bibfnamefont{A.~I.}
  \bibnamefont{Neishtadt}}, \emph{\bibinfo{title}{{Mathematical Aspects of
  Classical and Celestial Mechanics}}} (\bibinfo{publisher}{Springer},
  \bibinfo{address}{New York}, \bibinfo{year}{2006}).

\bibitem{Pechukas81}
\bibinfo{author}{\bibfnamefont{P.}~\bibnamefont{Pechukas}},
  \bibinfo{journal}{Ann. Rev. Phys. Chem.} \textbf{\bibinfo{volume}{32}},
  \bibinfo{pages}{159} (\bibinfo{year}{1981}).

\bibitem{Pechukas82}
\bibinfo{author}{\bibfnamefont{P.}~\bibnamefont{Pechukas}},
  \bibinfo{journal}{Ber. Buns. Ges.} \textbf{\bibinfo{volume}{86}},
  \bibinfo{pages}{372} (\bibinfo{year}{1982}).

\bibitem{Pollak85}
\bibinfo{author}{\bibfnamefont{E.}~\bibnamefont{Pollak}},
  \emph{\bibinfo{title}{{Periodic orbits and the theory of reactive
  scattering}}} (\bibinfo{publisher}{CRC Press}, \bibinfo{address}{Boca Raton},
  \bibinfo{year}{1985}), vol.~\bibinfo{volume}{3} of
  \emph{\bibinfo{series}{Theory of Chemical Reaction Dynamics}}, pp.
  \bibinfo{pages}{123--246}.

\bibitem{Wigner38}
\bibinfo{author}{\bibfnamefont{E.~P.} \bibnamefont{Wigner}},
  \bibinfo{journal}{Trans. Faraday Soc.} \textbf{\bibinfo{volume}{34}},
  \bibinfo{pages}{29} (\bibinfo{year}{1938}).

\bibitem{Truhlar96}
\bibinfo{author}{\bibfnamefont{D.~G.} \bibnamefont{Truhlar}},
  \bibinfo{author}{\bibfnamefont{B.~C.} \bibnamefont{Garrett}},
  \bibnamefont{and} \bibinfo{author}{\bibfnamefont{S.~J.}
  \bibnamefont{Klippenstein}}, \bibinfo{journal}{J. Phys. Chem.}
  \textbf{\bibinfo{volume}{100}}, \bibinfo{pages}{12711}
  (\bibinfo{year}{1996}).

\bibitem{Jaffe00}
\bibinfo{author}{\bibfnamefont{C.}~\bibnamefont{Jaffe}},
  \bibinfo{author}{\bibfnamefont{D.}~\bibnamefont{Farrelly}}, \bibnamefont{and}
  \bibinfo{author}{\bibfnamefont{T.}~\bibnamefont{Uzer}},
  \bibinfo{journal}{Phys. Rev. Lett.} \textbf{\bibinfo{volume}{84}},
  \bibinfo{pages}{610} (\bibinfo{year}{2000}).

\bibitem{Chandler78}
\bibinfo{author}{\bibfnamefont{D.}~\bibnamefont{Chandler}},
  \bibinfo{journal}{J. Chem. Phys.} \textbf{\bibinfo{volume}{68}},
  \bibinfo{pages}{2959} (\bibinfo{year}{1978}).

\bibitem{Chandler87}
\bibinfo{author}{\bibfnamefont{D.}~\bibnamefont{Chandler}},
  \emph{\bibinfo{title}{{Introduction to Modern Statistical Mechanics}}}
  (\bibinfo{publisher}{Oxford University Press}, \bibinfo{address}{New York},
  \bibinfo{year}{1987}).

\bibitem{DeLeon81}
\bibinfo{author}{\bibfnamefont{N.}~\bibnamefont{DeLeon}} \bibnamefont{and}
  \bibinfo{author}{\bibfnamefont{B.~J.} \bibnamefont{Berne}},
  \bibinfo{journal}{J. Chem. Phys.} \textbf{\bibinfo{volume}{75}},
  \bibinfo{pages}{3495} (\bibinfo{year}{1981}).

\bibitem{Berne82}
\bibinfo{author}{\bibfnamefont{B.~J.} \bibnamefont{Berne}},
  \bibinfo{author}{\bibfnamefont{N.}~\bibnamefont{DeLeon}}, \bibnamefont{and}
  \bibinfo{author}{\bibfnamefont{R.~O.} \bibnamefont{Rosenberg}},
  \bibinfo{journal}{J. Phys. Chem.} \textbf{\bibinfo{volume}{86}},
  \bibinfo{pages}{2166} (\bibinfo{year}{1982}).

\bibitem{Davis92}
\bibinfo{author}{\bibfnamefont{M.~J.} \bibnamefont{Davis}} \bibnamefont{and}
  \bibinfo{author}{\bibfnamefont{R.~T.} \bibnamefont{Skodje}},
  \emph{\bibinfo{title}{{Chemical Reactions as Problems in Nonlinear
  Dynamics}}} (\bibinfo{publisher}{JAI Press}, \bibinfo{address}{Greenwich,
  CT}, \bibinfo{year}{1992}), vol.~\bibinfo{volume}{3} of
  \emph{\bibinfo{series}{Advances in Classical Trajectory Methods}}, pp.
  \bibinfo{pages}{77--164}.

\bibitem{Bensimon84}
\bibinfo{author}{\bibfnamefont{D.}~\bibnamefont{Bensimon}} \bibnamefont{and}
  \bibinfo{author}{\bibfnamefont{L.}~\bibnamefont{Kadanoff}},
  \bibinfo{journal}{Physica D} \textbf{\bibinfo{volume}{13}},
  \bibinfo{pages}{82} (\bibinfo{year}{1984}).

\bibitem{Mackay84}
\bibinfo{author}{\bibfnamefont{R.~S.} \bibnamefont{MacKay}},
  \bibinfo{author}{\bibfnamefont{J.~D.} \bibnamefont{Meiss}}, \bibnamefont{and}
  \bibinfo{author}{\bibfnamefont{I.~C.} \bibnamefont{Percival}},
  \bibinfo{journal}{Physica D} \textbf{\bibinfo{volume}{13}},
  \bibinfo{pages}{55} (\bibinfo{year}{1984}).

\bibitem{Davis85}
\bibinfo{author}{\bibfnamefont{M.~J.} \bibnamefont{Davis}},
  \bibinfo{journal}{J. Chem. Phys.} \textbf{\bibinfo{volume}{83}},
  \bibinfo{pages}{1016} (\bibinfo{year}{1985}).

\bibitem{Channon80}
\bibinfo{author}{\bibfnamefont{S.~R.} \bibnamefont{Channon}} \bibnamefont{and}
  \bibinfo{author}{\bibfnamefont{J.~L.} \bibnamefont{Lebowitz}},
  \bibinfo{journal}{Ann. NY Acad. Sci.} \textbf{\bibinfo{volume}{357}},
  \bibinfo{pages}{108} (\bibinfo{year}{1980}).

\bibitem{Gray86a}
\bibinfo{author}{\bibfnamefont{S.~K.} \bibnamefont{Gray}} \bibnamefont{and}
  \bibinfo{author}{\bibfnamefont{M.~J.} \bibnamefont{Davis}},
  \bibinfo{journal}{J. Chem. Phys.} \textbf{\bibinfo{volume}{84}},
  \bibinfo{pages}{5389} (\bibinfo{year}{1986}).

\bibitem{Gray87}
\bibinfo{author}{\bibfnamefont{S.~K.} \bibnamefont{Gray}} \bibnamefont{and}
  \bibinfo{author}{\bibfnamefont{S.~A.} \bibnamefont{Rice}},
  \bibinfo{journal}{J. Chem. Phys.} \textbf{\bibinfo{volume}{86}},
  \bibinfo{pages}{2020} (\bibinfo{year}{1987}).

\bibitem{Marston89}
\bibinfo{author}{\bibfnamefont{C.~C.} \bibnamefont{Marston}} \bibnamefont{and}
  \bibinfo{author}{\bibfnamefont{N.}~\bibnamefont{DeLeon}},
  \bibinfo{journal}{J. Chem. Phys.} \textbf{\bibinfo{volume}{91}},
  \bibinfo{pages}{3392} (\bibinfo{year}{1989}).

\bibitem{DeAlmeida90}
\bibinfo{author}{\bibfnamefont{A.~M.~O.} \bibnamefont{DeAlmeida}},
  \bibinfo{author}{\bibfnamefont{N.}~\bibnamefont{DeLeon}},
  \bibinfo{author}{\bibfnamefont{M.~A.} \bibnamefont{Mehta}}, \bibnamefont{and}
  \bibinfo{author}{\bibfnamefont{C.~C.} \bibnamefont{Marston}},
  \bibinfo{journal}{Physica D} \textbf{\bibinfo{volume}{46}},
  \bibinfo{pages}{265} (\bibinfo{year}{1990}).

\bibitem{DeLeon91}
\bibinfo{author}{\bibfnamefont{N.}~\bibnamefont{DeLeon}},
  \bibinfo{author}{\bibfnamefont{M.~A.} \bibnamefont{Mehta}}, \bibnamefont{and}
  \bibinfo{author}{\bibfnamefont{R.~Q.} \bibnamefont{Topper}},
  \bibinfo{journal}{J. Chem. Phys.} \textbf{\bibinfo{volume}{94}},
  \bibinfo{pages}{8310} (\bibinfo{year}{1991}).

\bibitem{DeLeon91a}
\bibinfo{author}{\bibfnamefont{N.}~\bibnamefont{DeLeon}},
  \bibinfo{author}{\bibfnamefont{M.~A.} \bibnamefont{Mehta}}, \bibnamefont{and}
  \bibinfo{author}{\bibfnamefont{R.~Q.} \bibnamefont{Topper}},
  \bibinfo{journal}{J. Chem. Phys.} \textbf{\bibinfo{volume}{94}},
  \bibinfo{pages}{8329} (\bibinfo{year}{1991}).

\bibitem{DeLeon92}
\bibinfo{author}{\bibfnamefont{N.}~\bibnamefont{DeLeon}},
  \bibinfo{journal}{Chem. Phys. Lett.} \textbf{\bibinfo{volume}{189}},
  \bibinfo{pages}{371} (\bibinfo{year}{1992}).

\bibitem{DeLeon92a}
\bibinfo{author}{\bibfnamefont{N.}~\bibnamefont{DeLeon}}, \bibinfo{journal}{J.
  Chem. Phys.} \textbf{\bibinfo{volume}{96}}, \bibinfo{pages}{285}
  (\bibinfo{year}{1992}).

\bibitem{DeLeon94}
\bibinfo{author}{\bibfnamefont{N.}~\bibnamefont{DeLeon}} \bibnamefont{and}
  \bibinfo{author}{\bibfnamefont{S.}~\bibnamefont{Ling}}, \bibinfo{journal}{J.
  Chem. Phys.} \textbf{\bibinfo{volume}{101}}, \bibinfo{pages}{4790}
  (\bibinfo{year}{1994}).

\bibitem{Tersigni88}
\bibinfo{author}{\bibfnamefont{S.~H.} \bibnamefont{Tersigni}} \bibnamefont{and}
  \bibinfo{author}{\bibfnamefont{S.~A.} \bibnamefont{Rice}},
  \bibinfo{journal}{Berichte der Bunsen-Gesellschaft (Phys. Chem. Chem. Phys.)}
  \textbf{\bibinfo{volume}{92}}, \bibinfo{pages}{227} (\bibinfo{year}{1988}).

\bibitem{Tersigni90}
\bibinfo{author}{\bibfnamefont{S.~H.} \bibnamefont{Tersigni}},
  \bibinfo{author}{\bibfnamefont{P.}~\bibnamefont{Gaspard}}, \bibnamefont{and}
  \bibinfo{author}{\bibfnamefont{S.~A.} \bibnamefont{Rice}},
  \bibinfo{journal}{J. Chem. Phys.} \textbf{\bibinfo{volume}{92}},
  \bibinfo{pages}{1775} (\bibinfo{year}{1990}).

\bibitem{Gray86}
\bibinfo{author}{\bibfnamefont{S.~K.} \bibnamefont{Gray}},
  \bibinfo{author}{\bibfnamefont{S.~A.} \bibnamefont{Rice}}, \bibnamefont{and}
  \bibinfo{author}{\bibfnamefont{M.~J.} \bibnamefont{Davis}},
  \bibinfo{journal}{J. Phys. Chem.} \textbf{\bibinfo{volume}{90}},
  \bibinfo{pages}{3470} (\bibinfo{year}{1986}).

\bibitem{Zhao92}
\bibinfo{author}{\bibfnamefont{M.}~\bibnamefont{Zhao}} \bibnamefont{and}
  \bibinfo{author}{\bibfnamefont{S.~A.} \bibnamefont{Rice}},
  \bibinfo{journal}{J. Chem. Phys.} \textbf{\bibinfo{volume}{96}},
  \bibinfo{pages}{3542} (\bibinfo{year}{1992}).

\bibitem{Zhao92a}
\bibinfo{author}{\bibfnamefont{M.}~\bibnamefont{Zhao}} \bibnamefont{and}
  \bibinfo{author}{\bibfnamefont{S.~A.} \bibnamefont{Rice}},
  \bibinfo{journal}{J. Chem. Phys.} \textbf{\bibinfo{volume}{96}},
  \bibinfo{pages}{6654} (\bibinfo{year}{1992}).

\bibitem{Jang93}
\bibinfo{author}{\bibfnamefont{S.~M.} \bibnamefont{Jang}} \bibnamefont{and}
  \bibinfo{author}{\bibfnamefont{S.~A.} \bibnamefont{Rice}},
  \bibinfo{journal}{J. Chem. Phys.} \textbf{\bibinfo{volume}{99}},
  \bibinfo{pages}{9585} (\bibinfo{year}{1993}).

\bibitem{Rice96}
\bibinfo{author}{\bibfnamefont{S.~A.} \bibnamefont{Rice}} \bibnamefont{and}
  \bibinfo{author}{\bibfnamefont{M.~S.} \bibnamefont{Zhao}},
  \bibinfo{journal}{Int. J. Quantum Chem.} \textbf{\bibinfo{volume}{58}},
  \bibinfo{pages}{593} (\bibinfo{year}{1996}).

\bibitem{Toda02}
\bibinfo{author}{\bibfnamefont{M.}~\bibnamefont{Toda}}, \bibinfo{journal}{Adv.
  Chem. Phys.} \textbf{\bibinfo{volume}{123}}, \bibinfo{pages}{153}
  (\bibinfo{year}{2002}).

\bibitem{Arnold64}
\bibinfo{author}{\bibfnamefont{V.~I.} \bibnamefont{Arnold}},
  \bibinfo{journal}{Soviet Math. Doklady} \textbf{\bibinfo{volume}{5}},
  \bibinfo{pages}{581} (\bibinfo{year}{1964}).

\bibitem{Lochak99}
\bibinfo{author}{\bibfnamefont{P.}~\bibnamefont{Lochak}}, in
  \emph{\bibinfo{booktitle}{{Hamiltonian Systems with Three or More Degrees of
  Freedom}}}, edited by \bibinfo{editor}{\bibfnamefont{C.}~\bibnamefont{Simo}}
  (\bibinfo{publisher}{Kluwer}, \bibinfo{address}{Dordrecht},
  \bibinfo{year}{1999}), pp. \bibinfo{pages}{168--183}.

\bibitem{Heller95}
\bibinfo{author}{\bibfnamefont{E.~J.} \bibnamefont{Heller}},
  \bibinfo{journal}{J. Phys. Chem.} \textbf{\bibinfo{volume}{99}},
  \bibinfo{pages}{2625} (\bibinfo{year}{1995}).

\bibitem{Heller99}
\bibinfo{author}{\bibfnamefont{E.~J.} \bibnamefont{Heller}},
  \bibinfo{journal}{J. Phys. Chem. A} \textbf{\bibinfo{volume}{103}},
  \bibinfo{pages}{10433} (\bibinfo{year}{1999}).

\bibitem{Leitner97}
\bibinfo{author}{\bibfnamefont{D.~M.} \bibnamefont{Leitner}} \bibnamefont{and}
  \bibinfo{author}{\bibfnamefont{P.~G.} \bibnamefont{Wolynes}},
  \bibinfo{journal}{Phys. Rev. Lett.} \textbf{\bibinfo{volume}{79}},
  \bibinfo{pages}{55} (\bibinfo{year}{1997}).

\bibitem{Keshavamurthy07}
\bibinfo{author}{\bibfnamefont{S.}~\bibnamefont{Keshavamurthy}},
  \bibinfo{journal}{Int. Rev. Phys. Chem.} \textbf{\bibinfo{volume}{26}},
  \bibinfo{pages}{521} (\bibinfo{year}{2007}).

\bibitem{Toda05}
\bibinfo{author}{\bibfnamefont{M.}~\bibnamefont{Toda}}, \bibinfo{journal}{Adv.
  Chem. Phys.} \textbf{\bibinfo{volume}{130 A}}, \bibinfo{pages}{337}
  (\bibinfo{year}{2005}).

\bibitem{Shojiguchi08}
\bibinfo{author}{\bibfnamefont{A.}~\bibnamefont{Shojiguchi}},
  \bibinfo{author}{\bibfnamefont{C.~B.} \bibnamefont{Li}},
  \bibinfo{author}{\bibfnamefont{T.}~\bibnamefont{Komatsuzaki}},
  \bibnamefont{and} \bibinfo{author}{\bibfnamefont{M.}~\bibnamefont{Toda}},
  \bibinfo{journal}{Comm. Nonlinear Sci. Numerical Simulation}
  \textbf{\bibinfo{volume}{13}}, \bibinfo{pages}{857} (\bibinfo{year}{2008}).

\bibitem{Schofield94}
\bibinfo{author}{\bibfnamefont{S.~A.} \bibnamefont{Schofield}}
  \bibnamefont{and} \bibinfo{author}{\bibfnamefont{P.~G.}
  \bibnamefont{Wolynes}}, \bibinfo{journal}{Chem. Phys. Lett.}
  \textbf{\bibinfo{volume}{217}}, \bibinfo{pages}{497} (\bibinfo{year}{1994}).

\bibitem{Gruebele00}
\bibinfo{author}{\bibfnamefont{M.}~\bibnamefont{Gruebele}},
  \bibinfo{journal}{Adv. Chem. Phys.} \textbf{\bibinfo{volume}{114}},
  \bibinfo{pages}{193} (\bibinfo{year}{2000}).

\bibitem{Gruebele04}
\bibinfo{author}{\bibfnamefont{M.}~\bibnamefont{Gruebele}} \bibnamefont{and}
  \bibinfo{author}{\bibfnamefont{P.~G.} \bibnamefont{Wolynes}},
  \bibinfo{journal}{Acc. Chem. Res.} \textbf{\bibinfo{volume}{37}},
  \bibinfo{pages}{261} (\bibinfo{year}{2004}).

\bibitem{Leitner06}
\bibinfo{author}{\bibfnamefont{D.~M.} \bibnamefont{Leitner}} \bibnamefont{and}
  \bibinfo{author}{\bibfnamefont{P.~G.} \bibnamefont{Wolynes}},
  \bibinfo{journal}{Chem. Phys.} \textbf{\bibinfo{volume}{329}},
  \bibinfo{pages}{163} (\bibinfo{year}{2006}).

\bibitem{Leitner08}
\bibinfo{author}{\bibfnamefont{D.~M.} \bibnamefont{Leitner}} \bibnamefont{and}
  \bibinfo{author}{\bibfnamefont{M.}~\bibnamefont{Gruebele}},
  \bibinfo{journal}{Mol. Phys.} \textbf{\bibinfo{volume}{106}},
  \bibinfo{pages}{433} (\bibinfo{year}{2008}).

\bibitem{Schofield93}
\bibinfo{author}{\bibfnamefont{S.~A.} \bibnamefont{Schofield}}
  \bibnamefont{and} \bibinfo{author}{\bibfnamefont{P.~G.}
  \bibnamefont{Wolynes}}, \bibinfo{journal}{J. Chem. Phys.}
  \textbf{\bibinfo{volume}{98}}, \bibinfo{pages}{1123} (\bibinfo{year}{1993}).

\bibitem{Schofield95}
\bibinfo{author}{\bibfnamefont{S.~A.} \bibnamefont{Schofield}},
  \bibinfo{author}{\bibfnamefont{P.~G.} \bibnamefont{Wolynes}},
  \bibnamefont{and} \bibinfo{author}{\bibfnamefont{R.~E.} \bibnamefont{Wyatt}},
  \bibinfo{journal}{Phys. Rev. Lett.} \textbf{\bibinfo{volume}{74}},
  \bibinfo{pages}{3720} (\bibinfo{year}{1995}).

\bibitem{Keshavamurthy99}
\bibinfo{author}{\bibfnamefont{S.}~\bibnamefont{Keshavamurthy}},
  \bibinfo{journal}{Chem. Phys. Lett.} \textbf{\bibinfo{volume}{300}},
  \bibinfo{pages}{281} (\bibinfo{year}{1999}).

\bibitem{Zaslavsky05}
\bibinfo{author}{\bibfnamefont{G.}~\bibnamefont{Zaslavsky}},
  \emph{\bibinfo{title}{{Hamiltonian Chaos and Fractional Dynamics}}}
  (\bibinfo{publisher}{Oxford University Press}, \bibinfo{address}{New York},
  \bibinfo{year}{2005}).

\bibitem{Shojiguchi07}
\bibinfo{author}{\bibfnamefont{A.}~\bibnamefont{Shojiguchi}},
  \bibinfo{author}{\bibfnamefont{C.~B.} \bibnamefont{Li}},
  \bibinfo{author}{\bibfnamefont{T.}~\bibnamefont{Komatsuzaki}},
  \bibnamefont{and} \bibinfo{author}{\bibfnamefont{M.}~\bibnamefont{Toda}},
  \bibinfo{journal}{Phys. Rev. E} \textbf{\bibinfo{volume}{75}},
  \bibinfo{pages}{035204} (\bibinfo{year}{2007}).

\bibitem{Shojiguchi07a}
\bibinfo{author}{\bibfnamefont{A.}~\bibnamefont{Shojiguchi}},
  \bibinfo{author}{\bibfnamefont{C.~B.} \bibnamefont{Li}},
  \bibinfo{author}{\bibfnamefont{T.}~\bibnamefont{Komatsuzaki}},
  \bibnamefont{and} \bibinfo{author}{\bibfnamefont{M.}~\bibnamefont{Toda}},
  \bibinfo{journal}{Phys. Rev. E} \textbf{\bibinfo{volume}{76}},
  \bibinfo{pages}{056205} (\bibinfo{year}{2007}).

\bibitem{Wiggins90}
\bibinfo{author}{\bibfnamefont{S.}~\bibnamefont{Wiggins}},
  \bibinfo{journal}{Physica D} \textbf{\bibinfo{volume}{44}},
  \bibinfo{pages}{471} (\bibinfo{year}{1990}).

\bibitem{Wiggins94}
\bibinfo{author}{\bibfnamefont{S.}~\bibnamefont{Wiggins}},
  \emph{\bibinfo{title}{Normally hyperbolic invariant manifolds in dynamical
  systems}} (\bibinfo{publisher}{Springer-Verlag}, \bibinfo{year}{1994}).

\bibitem{Pollak78}
\bibinfo{author}{\bibfnamefont{E.}~\bibnamefont{Pollak}} \bibnamefont{and}
  \bibinfo{author}{\bibfnamefont{P.}~\bibnamefont{Pechukas}},
  \bibinfo{journal}{J. Chem. Phys.} \textbf{\bibinfo{volume}{69}},
  \bibinfo{pages}{1218} (\bibinfo{year}{1978}).

\bibitem{footnote1}
\bibinfo{note}{{The co-dimension of a submanifold is the dimension of the space
  in which the submanifold exists, minus the dimension of the submanifold. The
  significance of a submanifold being ``co-dimension one'' is that it is one
  less dimension than the space in which it exists. Therefore it can ``divide''
  the space and act as a separatrix, or barrier, to transport.}}

\bibitem{wwju}
\bibinfo{author}{\bibfnamefont{S.}~\bibnamefont{Wiggins}},
  \bibinfo{author}{\bibfnamefont{L.}~\bibnamefont{Wiesenfeld}},
  \bibinfo{author}{\bibfnamefont{C.}~\bibnamefont{Jaffe}}, \bibnamefont{and}
  \bibinfo{author}{\bibfnamefont{T.}~\bibnamefont{Uzer}},
  \bibinfo{journal}{Phys. Rev. Lett.} \textbf{\bibinfo{volume}{86(24)}},
  \bibinfo{pages}{5478} (\bibinfo{year}{2001}).

\bibitem{ujpyw}
\bibinfo{author}{\bibfnamefont{T.}~\bibnamefont{Uzer}},
  \bibinfo{author}{\bibfnamefont{C.}~\bibnamefont{Jaffe}},
  \bibinfo{author}{\bibfnamefont{J.}~\bibnamefont{Palacian}},
  \bibinfo{author}{\bibfnamefont{P.}~\bibnamefont{Yanguas}}, \bibnamefont{and}
  \bibinfo{author}{\bibfnamefont{S.}~\bibnamefont{Wiggins}},
  \bibinfo{journal}{Nonlinearity} \textbf{\bibinfo{volume}{15}},
  \bibinfo{pages}{957} (\bibinfo{year}{2002}).

\bibitem{Gillilan90}
\bibinfo{author}{\bibfnamefont{R.~E.} \bibnamefont{Gillilan}},
  \bibinfo{journal}{J. Chem. Phys.} \textbf{\bibinfo{volume}{93}},
  \bibinfo{pages}{5300} (\bibinfo{year}{1990}).

\bibitem{Gillilan91}
\bibinfo{author}{\bibfnamefont{R.~E.} \bibnamefont{Gillilan}} \bibnamefont{and}
  \bibinfo{author}{\bibfnamefont{G.~S.} \bibnamefont{Ezra}},
  \bibinfo{journal}{J. Chem. Phys.} \textbf{\bibinfo{volume}{94}},
  \bibinfo{pages}{2648} (\bibinfo{year}{1991}).

\bibitem{Toda95}
\bibinfo{author}{\bibfnamefont{M.}~\bibnamefont{Toda}}, \bibinfo{journal}{Phys.
  Rev. Lett.} \textbf{\bibinfo{volume}{74}}, \bibinfo{pages}{2670}
  (\bibinfo{year}{1995}).

\bibitem{WaalkensBurbanksWiggins04}
\bibinfo{author}{\bibfnamefont{H.}~\bibnamefont{Waalkens}},
  \bibinfo{author}{\bibfnamefont{A.}~\bibnamefont{Burbanks}}, \bibnamefont{and}
  \bibinfo{author}{\bibfnamefont{S.}~\bibnamefont{Wiggins}},
  \bibinfo{journal}{J. Phys. A} \textbf{\bibinfo{volume}{37}},
  \bibinfo{pages}{L257} (\bibinfo{year}{2004}).

\bibitem{WaalkensWiggins04}
\bibinfo{author}{\bibfnamefont{H.}~\bibnamefont{Waalkens}} \bibnamefont{and}
  \bibinfo{author}{\bibfnamefont{S.}~\bibnamefont{Wiggins}},
  \bibinfo{journal}{J. Phys. A} \textbf{\bibinfo{volume}{37}},
  \bibinfo{pages}{L435} (\bibinfo{year}{2004}).

\bibitem{WaalkensBurbanksWigginsb04}
\bibinfo{author}{\bibfnamefont{H.}~\bibnamefont{Waalkens}},
  \bibinfo{author}{\bibfnamefont{A.}~\bibnamefont{Burbanks}}, \bibnamefont{and}
  \bibinfo{author}{\bibfnamefont{S.}~\bibnamefont{Wiggins}},
  \bibinfo{journal}{J. Chem. Phys.}
  \textbf{\bibinfo{volume}{121}}(\bibinfo{number}{13}), \bibinfo{pages}{6207}
  (\bibinfo{year}{2004}).

\bibitem{WaalkensBurbanksWiggins05}
\bibinfo{author}{\bibfnamefont{H.}~\bibnamefont{Waalkens}},
  \bibinfo{author}{\bibfnamefont{A.}~\bibnamefont{Burbanks}}, \bibnamefont{and}
  \bibinfo{author}{\bibfnamefont{S.}~\bibnamefont{Wiggins}},
  \bibinfo{journal}{Physical Review Letters} \textbf{\bibinfo{volume}{95}},
  \bibinfo{pages}{084301} (\bibinfo{year}{2005}).

\bibitem{WaalkensBurbanksWiggins05c}
\bibinfo{author}{\bibfnamefont{H.}~\bibnamefont{Waalkens}},
  \bibinfo{author}{\bibfnamefont{A.}~\bibnamefont{Burbanks}}, \bibnamefont{and}
  \bibinfo{author}{\bibfnamefont{S.}~\bibnamefont{Wiggins}},
  \bibinfo{journal}{J. Phys. A} \textbf{\bibinfo{volume}{38}},
  \bibinfo{pages}{L759} (\bibinfo{year}{2005}).

\bibitem{SchubertWaalkensWiggins06}
\bibinfo{author}{\bibfnamefont{R.}~\bibnamefont{Schubert}},
  \bibinfo{author}{\bibfnamefont{H.}~\bibnamefont{Waalkens}}, \bibnamefont{and}
  \bibinfo{author}{\bibfnamefont{S.}~\bibnamefont{Wiggins}},
  \bibinfo{journal}{Phys. Rev. Lett.} \textbf{\bibinfo{volume}{96}},
  \bibinfo{pages}{218302} (\bibinfo{year}{2006}).

\bibitem{WaalkensSchubertWiggins08}
\bibinfo{author}{\bibfnamefont{H.}~\bibnamefont{Waalkens}},
  \bibinfo{author}{\bibfnamefont{R.}~\bibnamefont{Schubert}}, \bibnamefont{and}
  \bibinfo{author}{\bibfnamefont{S.}~\bibnamefont{Wiggins}},
  \bibinfo{journal}{Nonlinearity}
  \textbf{\bibinfo{volume}{21}}(\bibinfo{number}{1}), \bibinfo{pages}{R1}
  (\bibinfo{year}{2008}).

\bibitem{Komatsuzaki00}
\bibinfo{author}{\bibfnamefont{T.}~\bibnamefont{Komatsuzaki}} \bibnamefont{and}
  \bibinfo{author}{\bibfnamefont{R.~S.} \bibnamefont{Berry}},
  \bibinfo{journal}{J. Mol. Struct. THEOCHEM} \textbf{\bibinfo{volume}{506}},
  \bibinfo{pages}{55} (\bibinfo{year}{2000}).

\bibitem{Komatsuzaki02}
\bibinfo{author}{\bibfnamefont{T.}~\bibnamefont{Komatsuzaki}} \bibnamefont{and}
  \bibinfo{author}{\bibfnamefont{R.~S.} \bibnamefont{Berry}},
  \bibinfo{journal}{Adv. Chem. Phys.} \textbf{\bibinfo{volume}{123}},
  \bibinfo{pages}{79} (\bibinfo{year}{2002}).

\bibitem{Komatsuzaki05}
\bibinfo{author}{\bibfnamefont{T.}~\bibnamefont{Komatsuzaki}},
  \bibinfo{author}{\bibfnamefont{K.}~\bibnamefont{Hoshino}}, \bibnamefont{and}
  \bibinfo{author}{\bibfnamefont{Y.}~\bibnamefont{Matsunaga}},
  \bibinfo{journal}{Adv. Chem. Phys.} \textbf{\bibinfo{volume}{130 B}},
  \bibinfo{pages}{257} (\bibinfo{year}{2005}).

\bibitem{Wiesenfeld03}
\bibinfo{author}{\bibfnamefont{L.}~\bibnamefont{Wiesenfeld}},
  \bibinfo{author}{\bibfnamefont{A.}~\bibnamefont{Faure}}, \bibnamefont{and}
  \bibinfo{author}{\bibfnamefont{T.}~\bibnamefont{Johann}},
  \bibinfo{journal}{J. Phys. B} \textbf{\bibinfo{volume}{36}},
  \bibinfo{pages}{1319} (\bibinfo{year}{2003}).

\bibitem{Wiesenfeld04}
\bibinfo{author}{\bibfnamefont{L.}~\bibnamefont{Wiesenfeld}},
  \bibinfo{journal}{J. Phys. A} \textbf{\bibinfo{volume}{37}},
  \bibinfo{pages}{L143} (\bibinfo{year}{2004}).

\bibitem{Wiesenfeld04a}
\bibinfo{author}{\bibfnamefont{L.}~\bibnamefont{Wiesenfeld}},
  \bibinfo{journal}{Few Body Syst.} \textbf{\bibinfo{volume}{34}},
  \bibinfo{pages}{163} (\bibinfo{year}{2004}).

\bibitem{Gabern05}
\bibinfo{author}{\bibfnamefont{F.}~\bibnamefont{Gabern}},
  \bibinfo{author}{\bibfnamefont{W.~S.} \bibnamefont{Koon}},
  \bibinfo{author}{\bibfnamefont{J.~E.} \bibnamefont{Marsden}},
  \bibnamefont{and} \bibinfo{author}{\bibfnamefont{S.~D.} \bibnamefont{Ross}},
  \bibinfo{journal}{Physica D} \textbf{\bibinfo{volume}{211}},
  \bibinfo{pages}{391} (\bibinfo{year}{2005}).

\bibitem{Gabern06}
\bibinfo{author}{\bibfnamefont{F.}~\bibnamefont{Gabern}},
  \bibinfo{author}{\bibfnamefont{W.~S.} \bibnamefont{Koon}},
  \bibinfo{author}{\bibfnamefont{J.~E.} \bibnamefont{Marsden}},
  \bibnamefont{and} \bibinfo{author}{\bibfnamefont{S.~D.} \bibnamefont{Ross}},
  \bibinfo{journal}{Few-Body Systems} \textbf{\bibinfo{volume}{38}},
  \bibinfo{pages}{167} (\bibinfo{year}{2006}).

\bibitem{Gray80}
\bibinfo{author}{\bibfnamefont{S.~K.} \bibnamefont{Gray}},
  \bibinfo{author}{\bibfnamefont{W.~H.} \bibnamefont{Miller}},
  \bibinfo{author}{\bibfnamefont{Y.}~\bibnamefont{Yamaguchi}},
  \bibnamefont{and} \bibinfo{author}{\bibfnamefont{H.~F.}
  \bibnamefont{Schaefer}}, \bibinfo{journal}{J. Chem. Phys.}
  \textbf{\bibinfo{volume}{73}}, \bibinfo{pages}{2733} (\bibinfo{year}{1980}).

\bibitem{pmpb1}
\bibinfo{author}{\bibnamefont{{M Peri\'c and M. Mladenovi\'c and S. D.
  Peyerimhoff and R. J. Buenker}}}, \bibinfo{journal}{Chem. Phys.}
  \textbf{\bibinfo{volume}{82}}, \bibinfo{pages}{317} (\bibinfo{year}{1983}).

\bibitem{pmpb2}
\bibinfo{author}{\bibnamefont{{M Peri\'c and M. Mladenovi\'c and S. D.
  Peyerimhoff and R. J. Buenker}}}, \bibinfo{journal}{Chem. Phys.}
  \textbf{\bibinfo{volume}{86}}, \bibinfo{pages}{85} (\bibinfo{year}{1984}).

\bibitem{Lan93}
\bibinfo{author}{\bibfnamefont{B.~L.} \bibnamefont{Lan}} \bibnamefont{and}
  \bibinfo{author}{\bibfnamefont{J.~M.} \bibnamefont{Bowman}},
  \bibinfo{journal}{J. Phys. Chem.} \textbf{\bibinfo{volume}{97}},
  \bibinfo{pages}{12535} (\bibinfo{year}{1993}).

\bibitem{Bentley93}
\bibinfo{author}{\bibfnamefont{J.~A.} \bibnamefont{Bentley}},
  \bibinfo{author}{\bibfnamefont{C.~M.} \bibnamefont{Huang}}, \bibnamefont{and}
  \bibinfo{author}{\bibfnamefont{R.~E.} \bibnamefont{Wyatt}},
  \bibinfo{journal}{J. Chem. Phys.} \textbf{\bibinfo{volume}{98}},
  \bibinfo{pages}{5207} (\bibinfo{year}{1993}).

\bibitem{Tang94}
\bibinfo{author}{\bibfnamefont{H.}~\bibnamefont{Tang}},
  \bibinfo{author}{\bibfnamefont{S.}~\bibnamefont{Jang}},
  \bibinfo{author}{\bibfnamefont{M.}~\bibnamefont{Zhao}}, \bibnamefont{and}
  \bibinfo{author}{\bibfnamefont{S.~A.} \bibnamefont{Rice}},
  \bibinfo{journal}{J. Chem. Phys.}
  \textbf{\bibinfo{volume}{101}}(\bibinfo{number}{10}), \bibinfo{pages}{8737}
  (\bibinfo{year}{1994}).

\bibitem{Shah99}
\bibinfo{author}{\bibfnamefont{S.~P.} \bibnamefont{Shah}} \bibnamefont{and}
  \bibinfo{author}{\bibfnamefont{S.~A.} \bibnamefont{Rice}},
  \bibinfo{journal}{Faraday Disc.} \textbf{\bibinfo{volume}{113}},
  \bibinfo{pages}{319} (\bibinfo{year}{1999}).

\bibitem{Li05}
\bibinfo{author}{\bibfnamefont{C.~B.} \bibnamefont{Li}},
  \bibinfo{author}{\bibfnamefont{Y.}~\bibnamefont{Matsunaga}},
  \bibinfo{author}{\bibfnamefont{M.}~\bibnamefont{Toda}}, \bibnamefont{and}
  \bibinfo{author}{\bibfnamefont{T.}~\bibnamefont{Komatsuzaki}},
  \bibinfo{journal}{J. Chem. Phys.} \textbf{\bibinfo{volume}{123}},
  \bibinfo{pages}{Art. No. 184301} (\bibinfo{year}{2005}).

\bibitem{Gong05a}
\bibinfo{author}{\bibfnamefont{J.~B.} \bibnamefont{Gong}},
  \bibinfo{author}{\bibfnamefont{A.}~\bibnamefont{Ma}}, \bibnamefont{and}
  \bibinfo{author}{\bibfnamefont{S.~A.} \bibnamefont{Rice}},
  \bibinfo{journal}{J. Chem. Phys.} \textbf{\bibinfo{volume}{122}},
  \bibinfo{pages}{144311} (\bibinfo{year}{2005}).

\bibitem{Wigner39}
\bibinfo{author}{\bibfnamefont{E.~P.} \bibnamefont{Wigner}},
  \bibinfo{journal}{J. Chem. Phys.} \textbf{\bibinfo{volume}{7}},
  \bibinfo{pages}{646} (\bibinfo{year}{1939}).

\bibitem{Keck67}
\bibinfo{author}{\bibfnamefont{J.~C.} \bibnamefont{Keck}},
  \bibinfo{journal}{Adv. Chem. Phys.} \textbf{\bibinfo{volume}{XIII}},
  \bibinfo{pages}{85} (\bibinfo{year}{1967}).

\bibitem{Anderson95}
\bibinfo{author}{\bibfnamefont{J.~B.} \bibnamefont{Anderson}},
  \bibinfo{journal}{Adv. Chem. Phys.} \textbf{\bibinfo{volume}{XCI}},
  \bibinfo{pages}{381} (\bibinfo{year}{1995}).

\bibitem{Wales03}
\bibinfo{author}{\bibfnamefont{D.~J.} \bibnamefont{Wales}},
  \emph{\bibinfo{title}{Energy {L}andscapes}} (\bibinfo{publisher}{Cambridge
  University Press}, \bibinfo{address}{Cambridge}, \bibinfo{year}{2003}).

\bibitem{footnote0}
\bibinfo{note}{{Such a separation is assumed to be meaningful for the range of
  energies considered here.}}

\bibitem{Arnold78}
\bibinfo{author}{\bibfnamefont{V.~I.} \bibnamefont{Arnold}},
  \emph{\bibinfo{title}{Mathematical Methods of Classical Mechanics}},
  vol.~\bibinfo{volume}{60} of \emph{\bibinfo{series}{Graduate Texts in
  Mathematics}} (\bibinfo{publisher}{Springer}, \bibinfo{address}{New York,
  Heidelberg, Berlin}, \bibinfo{year}{1978}).

\bibitem{Binney85}
\bibinfo{author}{\bibfnamefont{J.}~\bibnamefont{Binney}},
  \bibinfo{author}{\bibfnamefont{O.~E.} \bibnamefont{Gerhard}},
  \bibnamefont{and} \bibinfo{author}{\bibfnamefont{P.}~\bibnamefont{Hut}},
  \bibinfo{journal}{Mon. Not. Roy. Astron. Soc.}
  \textbf{\bibinfo{volume}{215}}, \bibinfo{pages}{59} (\bibinfo{year}{1985}).

\bibitem{Meyer86}
\bibinfo{author}{\bibfnamefont{H.-D.} \bibnamefont{Meyer}},
  \bibinfo{journal}{J. Chem. Phys.} \textbf{\bibinfo{volume}{84}},
  \bibinfo{pages}{3147} (\bibinfo{year}{1986}).

\bibitem{Toller85}
\bibinfo{author}{\bibfnamefont{M.}~\bibnamefont{Toller}},
  \bibinfo{author}{\bibfnamefont{G.}~\bibnamefont{Jacucci}},
  \bibinfo{author}{\bibfnamefont{G.}~\bibnamefont{DeLorenzi}},
  \bibnamefont{and} \bibinfo{author}{\bibfnamefont{C.~P.} \bibnamefont{Flynn}},
  \bibinfo{journal}{Phys. Rev. B} \textbf{\bibinfo{volume}{32}},
  \bibinfo{pages}{2082} (\bibinfo{year}{1985}).

\bibitem{Mackay90}
\bibinfo{author}{\bibfnamefont{R.~S.} \bibnamefont{MacKay}},
  \bibinfo{journal}{Phys. Lett. A} \textbf{\bibinfo{volume}{145}},
  \bibinfo{pages}{425} (\bibinfo{year}{1990}).

\bibitem{Stember07}
\bibinfo{author}{\bibfnamefont{J.~N.} \bibnamefont{Stember}} \bibnamefont{and}
  \bibinfo{author}{\bibfnamefont{G.~S.} \bibnamefont{Ezra}},
  \bibinfo{journal}{Chem. Phys.} \textbf{\bibinfo{volume}{337}},
  \bibinfo{pages}{11} (\bibinfo{year}{2007}).

\bibitem{Li06}
\bibinfo{author}{\bibfnamefont{C.~B.} \bibnamefont{Li}},
  \bibinfo{author}{\bibfnamefont{A.}~\bibnamefont{Shojiguchi}},
  \bibinfo{author}{\bibfnamefont{M.}~\bibnamefont{Toda}}, \bibnamefont{and}
  \bibinfo{author}{\bibfnamefont{T.}~\bibnamefont{Komatsuzaki}},
  \bibinfo{journal}{Phys. Rev. Lett.} \textbf{\bibinfo{volume}{97}},
  \bibinfo{pages}{Art. No. 028302} (\bibinfo{year}{2006}).

\bibitem{MCH}
\bibinfo{author}{\bibfnamefont{J.~N.} \bibnamefont{Murrell}},
  \bibinfo{author}{\bibfnamefont{S.}~\bibnamefont{Carter}}, \bibnamefont{and}
  \bibinfo{author}{\bibfnamefont{L.~O.} \bibnamefont{Halonen}},
  \bibinfo{journal}{J. Mol. Spect.} \textbf{\bibinfo{volume}{93}},
  \bibinfo{pages}{307} (\bibinfo{year}{1982}).

\bibitem{footnote3}
\bibinfo{note}{{The quantity $f_{\text{T}}$ is therefore the effective value
  for the fraction of trapped states as determined on the timescale of the
  calculation. The true behavior of the survival probability $P_{\text{S}}(t)$
  at \emph{extremely} long times has not however been fully characterized, and
  merits further study.}}

\bibitem{Paskauskas08}
\bibinfo{author}{\bibfnamefont{R.}~\bibnamefont{Paskauskas}},
  \bibinfo{author}{\bibfnamefont{C.}~\bibnamefont{Chandre}}, \bibnamefont{and}
  \bibinfo{author}{\bibfnamefont{T.}~\bibnamefont{Uzer}},
  \bibinfo{journal}{Phys. Rev. Lett.} \textbf{\bibinfo{volume}{100}},
  \bibinfo{pages}{083001} (\bibinfo{year}{2008}).

\bibitem{Guckenheimer83}
\bibinfo{author}{\bibfnamefont{J.}~\bibnamefont{Guckenheimer}}
  \bibnamefont{and} \bibinfo{author}{\bibfnamefont{P.}~\bibnamefont{Holmes}},
  \emph{\bibinfo{title}{{Nonlinear Oscillations, Dynamical Systems, and
  Bifurcations of Vector Fields}}} (\bibinfo{publisher}{Springer},
  \bibinfo{address}{New York}, \bibinfo{year}{1983}).

\bibitem{Birkhoff31}
\bibinfo{author}{\bibfnamefont{G.~D.} \bibnamefont{Birkhoff}},
  \bibinfo{journal}{Bull. Am. Math. Soc.} \textbf{\bibinfo{volume}{38}},
  \bibinfo{pages}{361} (\bibinfo{year}{1931}).

\bibitem{Ott02}
\bibinfo{author}{\bibfnamefont{E.}~\bibnamefont{Ott}},
  \emph{\bibinfo{title}{{Chaos in Dynamical Systems}}}
  (\bibinfo{publisher}{Cambridge University Press},
  \bibinfo{address}{Cambridge}, \bibinfo{year}{2002}), 2nd ed.

\end{thebibliography}
\end{document}